\documentclass[aps,preprint,superscriptaddress]{revtex4-1}
\usepackage{amsmath}
\usepackage{bm}
\usepackage{graphicx}

\begin{document}

\title{Crystal Phases of Charged Interlayer Excitons\\ in van der Waals Heterostructures}

\author{Igor V.~Bondarev}
\email[Corresponding author email:~]{ibondarev@nccu.edu}
\affiliation{Department of Mathematics \& Physics, North Carolina Central University, Durham, NC 27707, USA}
\author{Oleg L.~Berman}
\author{Roman Ya.~Kezerashvili}
\affiliation{Physics Department, New York City College of Technology, City University of New York, NY 11201, USA}
\affiliation{Graduate School and University Center, City University of New York, NY 10016, USA}
\author{Yurii E.~Lozovik}
\affiliation{Institute of Spectroscopy, Russian Academy of Sciences, 142190 Troitsk, Moscow Region, Russia}
\affiliation{National Research University "Higher School of Economics",\vskip-0.05cm Tikhonov Moscow Institute of Electronics \& Mathematics, 123458 Moscow, Russia}

\begin{abstract}
Throughout the years, strongly correlated coherent states of excitons have been the subject of intense theoretical and experimental studies. This topic has recently boomed due to new emerging quantum materials such as van der Waals (vdW) bound atomically thin layers of transition metal dichalcogenides (TMDs). We analyze the collective properties of charged interlayer excitons observed recently in bilayer TMD heterostructures. We predict new strongly correlated phases --- crystal and Wigner crystal --- that can be selectively realized with TMD bilayers of properly chosen electron-hole effective masses by just varying their interlayer separation distance. Our results open up new avenues for nonlinear coherent control, charge transport and spinoptronics applications with quantum vdW heterostuctures.
\end{abstract}

\maketitle

Strongly correlated coherent states of excitons have been a subject of intense theoretical and experimental studies over the last decades~\cite{KeldyshKozlov68,LozovikYudson,Ogawa90,LozovikPRL07,Berman08,Kotthaus13,Kezer14,Fogler14,Suris16,JonFinley}. The topic has gained momentum recently due to new emerging materials of reduced dimensionality such as atomically thin van der Waals (vdW) bound layers of semiconducting transition metal dichalcogenides (TMDs)~\cite{MakShan16,APrev17,Wang18,Drummond18,Thygesen18,LozovikUFN18,Kezer19,Shklov19}.~These layered quasi-two-dimensional (2D) semiconductors make the exciton formation possible of electrons and holes located in distinct layers~\cite{Rivera2015,Ross2017,Baranowski2017,Miller17,Lius-PKim19,Geim20}. Due to the dimensionality reduction and because of a greatly reduced electron-hole wavefunction overlap, interlayer (or indirect) excitons thus formed have large binding energies and long lifetimes. Being electrically neutral, they feature a permanent electric dipole moment directed perpendicular to the layers, offering tunability of their quantum states by an external electric field. Similar to indirect excitons in conventional GaAs based coupled quantum well systems~\cite{Snoke,ButovJETP}, the interlayer excitons (IE) in vdW hetero\-structures can be coupled to light to form dipolar exciton-polaritons, allow\-ing control of quantum phenomena such as electromagnetically induced transparency, adiabatic photon-to-electron transfer, room-temperature Bose-Einstein condensation (BEC) and superconductivity~\cite{Li17,Szymanska2012,Cristofolini2012,Imamoglu16,Kavokin16,BondSnoke20}.

\begin{figure}[t]
\includegraphics[scale=0.5]{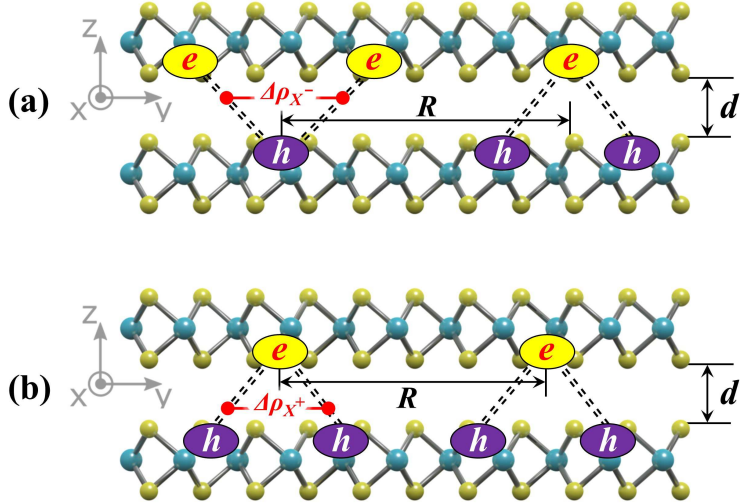}
\caption{The compound structure and pairwise interaction geometry for the unlike-charge~(a) and like-charge~(b) interlayer exciton complexes (trions) in a bilayer quasi-2D semiconductor.}
\label{fig1}
\end{figure}

For bilayer TMD heterostructures, controlled optical and electrical generation of IEs and \emph{charged} IEs (CIEs, also known as trions formed by indirect excitons~\cite{BondVlad18}) has lately been achieved~\cite{Lius-PKim19,Geim20}. Their in-plane propagation through the sample was adjusted by the excitation power and perpendicular electrostatic field. These experiments exhibit a unique potential of TMD bilayers for achieving precise control over compound quantum particles of both bosonic (IE) and fermionic (CIE) nature. The CIEs offer even more flexibility in this respect as they have both net charge and permanent dipole moment as well as non-zero spin (Fig.~\ref{fig1}), to allow for electrical tunability and optical spin manipulation in charge transport and spinoptronics experiments with quasi-2D vdW hetero\-structures.

Here, we consider the collective properties of the negative and positive CIEs starting with their binding energies in bilayer quasi-2D semiconductor heterostructures. We derive the general analytical expressions as functions of the electron-hole effective mass ratio and interlayer separation distance to explain the experimental evidence earlier reported for the negative CIE to have a greater binding energy than that of the positive CIE~\cite{Lius-PKim19}. Our analysis of the pairwise interactions between the CIEs, as sketched in Fig.~\ref{fig1}, exhibits two scenarios for crystallization phase transitions in the collective multi\-particle CIE system. They are the crystallization of the unlike-charge CIEs and the Wigner crystallization of the like-charge CIEs, which can be selectively realized in practice by choosing bilayers with appropriate electron-hole effective mass ratio and interlayer separation in addition to the standard technique of electrostatic doping. We conclude that this strongly correlated multiexciton phenomenon of CIE crystallization can be realized in layered van der Waals heterostructures such as double bilayer graphene and bilayer TMD systems~\cite{Li17,Lius-PKim19}, to open up new avenues for nonlinear coherent optical control and spinoptronics applications with charged interlayer excitons.

\subsection{The binding energy}

The compound structure of the CIE complexes of interest is sketched in Fig.~\ref{fig1}.~We use the configuration space approach~\cite{Bondarev2016} to derive the binding energy expressions for the CIEs as functions of their electron-hole effective mass ratio $\sigma\!=\!m_e/m_h$ and interlayer separation distance $d$. This approach was recently proven to be efficient as applied to quasi-1D~\cite{Bondarev11PRB} and quasi-2D bilayer semiconductors~\cite{BondVlad18} where it offers easily tractable analytical solutions to reveal universal relations between the binding energy of the complex of interest and that of the 1D-exciton or that of the indirect (interlayer) exciton~\cite{LeavittLittle}, respectively. The method itself was originally pioneered by Landau~\cite{LandauQM}, Gor'kov and Pitaevski~\cite{Pitaevski63}, Holstein and Herring~\cite{Herring} in their studies of molecular binding and magnetism.

The negative $X^{\!-\!}$ (positive $X^{+}$) trion complex in Fig.~\ref{fig1} can be viewed as two equivalent IEs sharing the same hole (electron). The CIE bound state then forms due to the exchange under-barrier tunneling between the \emph{equivalent} configurations of the electron-hole system in the configuration space of the \emph{two independent} relative electron-hole motion coordinates representing the two equivalent IEs that are separated by the center-of-mass-to-center-of-mass distance $\Delta\rho$. The binding strength is controlled by the exchange tunneling rate integral $J_{X^\pm}(\Delta\rho)$. The CIE binding energy is
\begin{equation}
E_{\!X^{^{\pm}}}(\sigma,d)=-J_{\!X^{^{\pm}}}(\Delta\rho\!=\!\Delta\rho_{X^{^{\!\pm}}})
\label{bindingE}
\end{equation}
with $\Delta\rho_{X^\pm}$ to be determined from an appropriate variational procedure to maximize the tunneling rate, which corresponds to the CIE ground state. This approach gives an upper bound for the (negative) \emph{ground} state binding energy of an exciton complex of interest~\cite{Bondarev2016,Bondarev11PRB,BondVlad18}. It captures essential kinematics of the formation of the complex and helps understand the general physical principles to underlie its stability.

Using the configuration space method for solving the CIE ground state binding energy problem, we obtain (see \emph{Methods})
\begin{eqnarray}
J_{\!X^{^{\pm}}}(\Delta\rho)=2N^4\Delta\rho^2\exp\!\left[-2\alpha\!\left(\!\sqrt{\Delta\rho^2+4d^2}-2d\right)\right]\hskip3.5cm\label{JXfin}\\
\times\left[\frac{\alpha}{\sqrt{\Delta\rho^2+4d^2}}+\!\frac{1}{2\big(r_0+\Big\{\!\!\begin{array}{c}1\\[-0.35cm]
\,\sigma\end{array}\!\!\Big\}\Delta\rho/\lambda\big)(\alpha\Delta\rho-1)}\right]\left(\frac{r_0+\Big\{\!\!\begin{array}{c}1\\[-0.35cm]
\,\sigma\end{array}\!\!\Big\}\Delta\rho/\lambda}{r_0+\Delta\rho}\right)^{\displaystyle\frac{\lambda\Delta\rho}{\Big\{\!\!\begin{array}{c}\sigma\\[-0.3cm]1\end{array}\!\!\Big\}
\left(\alpha\Delta\rho-1\right)}},\nonumber
\end{eqnarray}
where $\alpha\!=\!2/(1+2\sqrt{d}\,)$ and $N\!\!=\!4/\!\sqrt{1\!+4\sqrt{d}+8d(1\!+\!\sqrt{d}\,)}$ are the interlayer separation dependent constants coming from the indirect (interlayer) exciton wave function~\cite{LeavittLittle}, and the upper or lower term should be taken in the curly brackets for the positive or negative CIE, respectively. Here the 3D "atomic units"\space are used~\cite{LandauQM,Pitaevski63,Herring,LeavittLittle}, with distance and energy measured in the units of exciton Bohr radius $a^\ast_B\!\!=\!0.529\,\mbox{\AA}\,\varepsilon/\mu$ and exciton Rydberg energy $Ry^\ast\!\!=\!\hbar^2\!/(2\mu\,m_0a_B^{\ast2})\!=\!e^2\!/(2\varepsilon a_B^\ast)\!=\!13.6\,\mbox{eV}\,\mu/\varepsilon^2$, respectively, $\varepsilon$ represents the \emph{effective} average dielectric constant of the bilayer heterostructure and $\mu\!=\!m_e/(\lambda\,m_0)$ stands for the exciton reduced effective mass (in the units of free electron mass $m_0$) with $\lambda\!=\!1+m_e/m_h\!=\!1+\sigma$. The image-charge effects are neglected~\cite{LeavittLittle}. To properly take into account the screening effect for the charges forming the CIEs as sketched in Fig.~\ref{fig1}, we used the Keldysh-Rytova (KR) interaction potential energy (see Refs.~\cite{KeldyshRytova}) approximated by elementary functions in the form (atomic units)
\begin{eqnarray}
V_{\texttt{eff}}(\rho)=\frac{1}{r_0}\left[\ln\!\left(\!1+\frac{r_0}{\rho}\!\right)+(\ln2-\gamma)e^{-\rho/r_0}\right]
\label{keldyshpot}
\end{eqnarray}
proposed for atomically thin layers in Ref.~\cite{Rubio11}, to represent the effective electrostatic potential energy for like charges in monolayers. Here, $\rho$ is the in-plane intercharge distance and $r_0\!=\!2\pi\chi_\texttt{2D}$ is the screening length parameter with $\chi_\texttt{2D}$ being the in-plane polarizability of 2D material~\cite{Rubio11,Berkelbach2013}. For unlike charges the interlayer electrostatic potential energy is taken in the standard screened Coulomb form $V_{\texttt{C}}(r)\!=\!-1/r$ with $r\!=\!\sqrt{\rho^2+d^2}$ (atomic units).

The function $J_{\!X^{^{\pm}}}(\Delta\rho)$ in Eq.~(\ref{JXfin}) is clearly seen to have a maximum. It tends to become a negative when $\alpha\Delta\rho<1$ in the second term in the square brackets, which is always the case for large enough $d$ whereby $\alpha\approx1/\sqrt{d}\sim0$ and the first term in the square brackets is negligible, whereas for $\alpha\Delta\rho>1$ it is manifestly positive and approaching zero as $\Delta\rho$ increases. Extremum seeking under the condition that $\Delta\rho>1$ to \emph{only} include the leading terms in small $1/\Delta\rho$, gives a compact result (see \emph{Methods})
\begin{equation}
\Delta\rho_{\!X^{^{\pm}}}=\frac{7\alpha-1-\Big\{\!\!\begin{array}{c}\sigma\\[-0.25cm]1/\sigma\end{array}\!\!\Big\}}{2\alpha^2}
-\!\left(3+2\Big\{\!\!\begin{array}{c}\sigma\\[-0.25cm]1/\sigma\end{array}\!\!\Big\}\!\right)r_0\,.
\label{Drho0Xpm}
\end{equation}
Substituting this in Eq.~(\ref{bindingE}), one obtains the positive and negative CIE binding energies of interest.

\begin{figure}[t]
\includegraphics[scale=0.5]{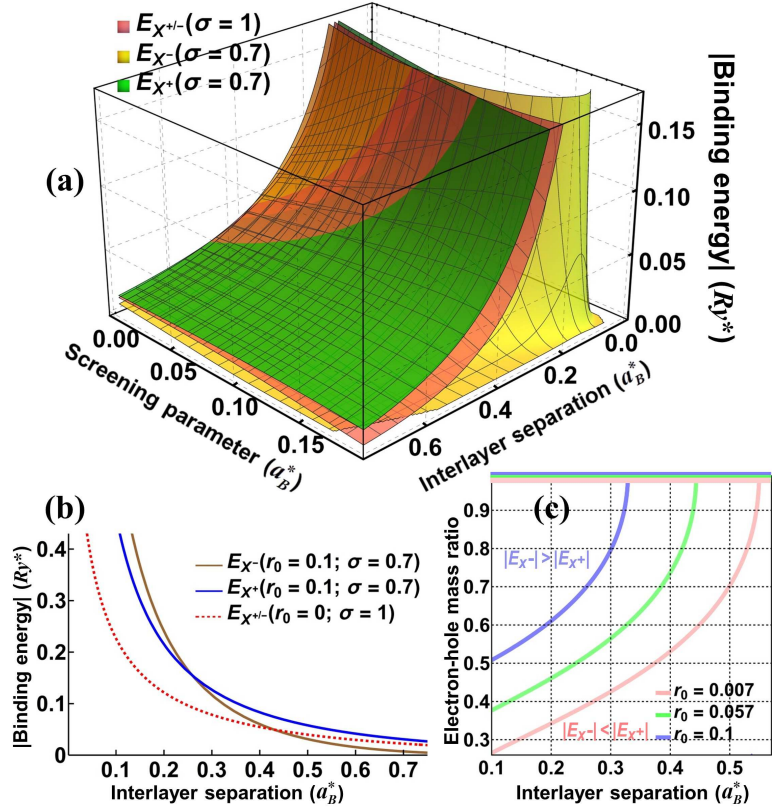}
\caption{(a)~Binding energies of the positive and negative CIEs as functions of the interlayer separation $d$ and screening length $r_0$ as given for $\sigma\!=\!1$ and $0.7$ by Eqs.~(\ref{bindingE}), (\ref{JXfin}) and (\ref{Drho0Xpm}). (b)~Crosscuts of (a) for $r_0\!=\!0$ and~$0.1$ to show the binding energy splitting for the positive and negative CIEs with unequal electron-hole masses. (c)~Solutions to equation $E_{\!X^{^{+\!}}}(\sigma,d)\!=\!E_{\!X^{^{-\!}}}(\sigma,d)$ for three values of the screening length.}
\label{fig2}
\end{figure}

Figure~\ref{fig2}~(a) shows the binding energies $E_{\!X^{^{+}}}$ and $E_{\!X^{^{-}}}$ calculated from Eqs.~(\ref{bindingE}), (\ref{JXfin}) and (\ref{Drho0Xpm}) with $\sigma\!=\!1$ and $0.7$ as functions of $d$ and $r_0$. For $\sigma\!=\!1$ they coincide~\cite{BondVlad18}. For $\sigma\!=\!0.7$ the positive-negative CIE binding energy splitting is seen to occur in the entire domain of parameters used. Figure~\ref{fig2}~(b) shows the crosscuts of Fig.~\ref{fig2}~(a) for $r_0\!=\!0$ and~$0.1$ to exhibit the remarkable features of the screening and binding energy splitting effects. The screening of like charges in the CIE complex is seen to increase its binding energy. The $X^{\pm}$ trion energy splitting at short $d$ is such that $|E_{\!X^{^{-\!}}}|\!>\!|E_{\!X^{^{+\!}}}|$, which agrees with and thus explains the measurements reported recently for (h-BN)-encapsulated MoSe$_2$--WSe$_2$ bilayer hetero\-structures~\cite{Lius-PKim19}. As $d$ increases the crossover occurs to give $|E_{\!X^{^{-\!}}}|\!<\!|E_{\!X^{^{+\!}}}|$ with $|E_{\!X^{^{-}}}|$ quickly going down to zero, which is also seen in Fig.~\ref{fig2}~(a). On closer inspection of Eqs.~(\ref{JXfin}) and (\ref{Drho0Xpm}) it can be seen though that $|E_{\!X^{^{-\!}}}|$ and $|E_{\!X^{^{+\!}}}|$ swap places for $\sigma\!>\!1$ (not shown here), thereby offering an extra functionality for properly fabricated vdW heterostructures~\cite{Larentis18,TMDmass18}.

Equation $E_{\!X^{^{+\!}}}(\sigma,d)\!=\!E_{\!X^{^{-\!}}}(\sigma,d)$ links the electron-hole mass ratio $\sigma$ and interlayer separation $d$ at which the crossover occurs. For $\sigma\!=\!1$ it turns into an identity~\cite{BondVlad18}. The three lines in Fig.~\ref{fig2}~(c) present the nontrivial solution to this equation, $\sigma(d)$, for three different $r_0$ values. The screening is seen to shrink the $|E_{\!X^{^{-\!}}}|\!>\!|E_{\!X^{^{+\!}}}|$ domain and expand the $|E_{\!X^{^{-\!}}}|\!<\!|E_{\!X^{^{+\!}}}|$ domain (above and below the solution line, respectively). Since the greater binding energy increases the formation probability, these domains are also those to preferentially form the $X^{\!-}$ and $X^{+}$ trion, respectively, while the constraint $E_{\!X^{^{-\!}}}\!=\!E_{\!X^{^{+\!}}}$ defines the line of equal $X^\pm$ formation probabilities. Thus by varying $d$ for a properly chosen TMD bilayer composition with known~$\sigma$, one can selectively control \emph{intrinsic} positive/negative CIE formation in an undoped heterostructure as opposed to the electrostatic doping technique.

\subsection{Unlike-charge trion crystallization}

For undoped structures of two monolayers with $\sigma\!=\!1$ as well as for those with $\sigma\!\ne\!1$ fabricated to hit the $E_{\!X^{^{-\!}}}\!=\!E_{\!X^{^{+\!}}}$ line, both $X^{\!-}$ and $X^{+\!}$ trions are equally likely to form under intense external irradiation at not too high temperatures $T\!<\!|E_{\!X^{^{\pm\!}}}|/k_B$. This results in an overall neutral two-component many-particle mixture of $X^{\!-}$ and $X^{+\!}$ trions. The aggregate state of a many-particle system is defined by its Helmholtz free energy consisting of the total energy term and the entropy term. The entropy term becomes dominant at high $T$ to favor configurations with greater randomness. At not too high $T$ the total energy term --- the sum of kinetic, potential and binding energies of individual particles --- overcomes the entropy term so that an ordered state is favored, with the order-disorder transition being predominantly determined by the interparticle pairwise interaction potential energy~\cite{Kubo}.

\begin{figure}[t]
\includegraphics[scale=0.5]{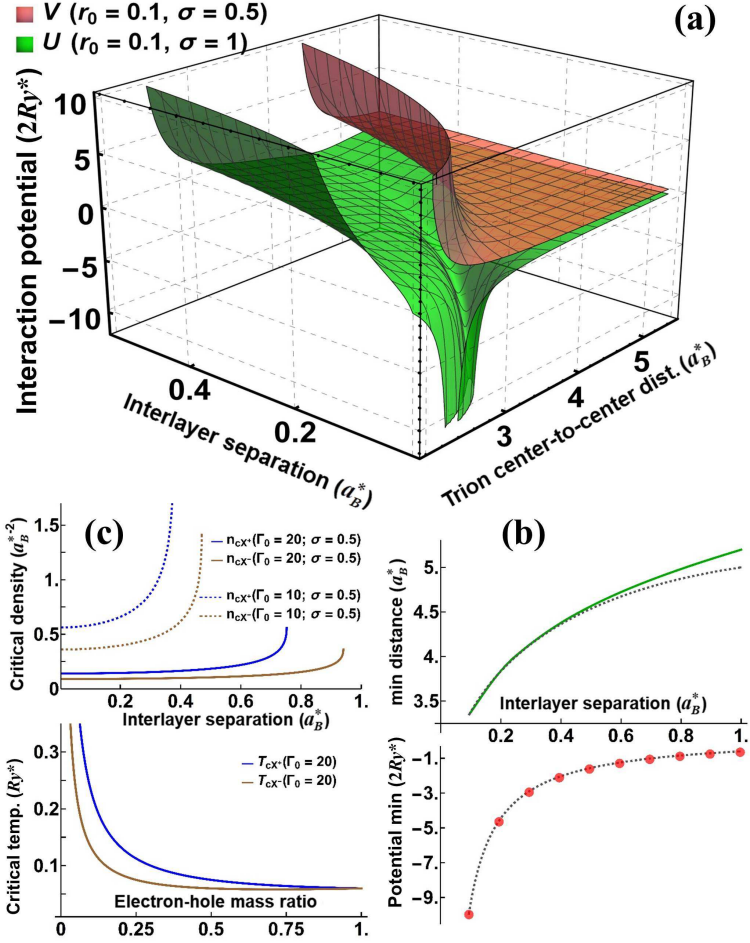}
\caption{(a)~The exact unlike-charge $U$ (attractive) and like-charge $V$ (repulsive) electrostatic interaction potentials calculated as functions of $R$ and $d$ for the trion pairs with $\sigma\!=\!1$~and~0.5, respectively, in their symmetry-promoted most likely configurations shown in Fig.~\ref{fig1}. ~(b)~The approximate analytical (dashed black lines) and calculated $d$-dependences of $R_\texttt{min}$ (top, green line) and $U_\texttt{min}$ (bottom, red dots) for the $U$-potential surface in (a). ~(c)~Critical densities (top) and temperatures (bottom) for the Wigner crystallization transition of the many-paticle like-charge trion system.}
\label{fig3}
\end{figure}

The long-range Coulomb interaction of the pair of CIEs (trions) is strengthened by their permanent dipole moments directed perpendicular to the plane of the structure. Their actual \emph{exact} interaction potential depends on the relative orientation of the triangles formed by the three charges in a trion complex. The exact potential includes nine terms to couple electrons and holes in two complexes by means of the $V_{\texttt{eff}}(R,\Delta\rho_{\!X^{^{\pm}}},r_0)$ and $V_{\texttt{C}}(R,\Delta\rho_{\!X^{^{\pm}}},d)$ potentials, where $R$ is the trion center-to-center distance (see \emph{Methods}). Figure~\ref{fig3}~(a) shows the exact interaction potentials $U$ and $V$ as functions of $R$ and $d$ for unlike- and like-charge trion pairs (shown for $\sigma\!=\!1$~and~0.5, respectively; no major change observed with the variation of~$\sigma$) in their symmetry-promoted most likely configurations presented in Fig.~\ref{fig1}. The unlike-charge trion pairwise interaction potential exhibits a deep attractive (negative) minimum and a strongly repulsive (positive) core for all $d$ in the range presented, in contrast with the manifestly repulsive like-charge trion pairwise potential. This is what makes the order-disorder transition in the two-component unlike-charge trion system identical to that in an $AB$ type alloy with $A$ and $B$ components randomly mixed at high $T$ and ordered on the ionic-crystal-type superlattice of interpenetrating $a$- and $b$-sublattices below $T^\texttt{(N)}_{c}\!=\!zv/2k_B$, the N\'{e}el temperature~\cite{Kubo}. Here, $z$ is the number of the nearest neighbors on the superlattice and $v\!=\!(v_{AA}\!+\!v_{BB})/2\!-\!v_{AB}\!\!>\!0$ is the combined nearest-neighbor coupling constant written in terms of those of respective sublattices. In our case here, the ordering below $T^\texttt{(N)}_{c}$ creates 1D chains ($z\!=\!2$) of the two interpenetrating sublattices with collinear CIE permanent dipole moments in each of the two. In full analogy, taking the parameters $R_\texttt{min}$ and $U_{\texttt{min}}\!=\!U(R_\texttt{min})$ of the minimum of the potential $U$ in Fig.~\ref{fig3}~(a) to represent the chain period and the unlike-charge trion coupling constant, respectively, one obtains $T^\texttt{(N)}_{c}\!\approx\![V(2R_\texttt{min})\!-\!U(R_\texttt{min})]/k_B\!\approx\!|U_{\texttt{min}}|/k_B$. Here, $V(2R_\texttt{min})\!\approx\!0$ stands for the repulsive interaction coupling constant of the like-charge trions whose sublattice period is twice greater than the period of the chain.

The top and bottom panels in Fig.~\ref{fig3}~(b) present the exact $d$-dependences of $R_\texttt{min}$ and $U_\texttt{min}$ calculated for the $U$-potential surface shown in Fig.~\ref{fig3}~(a). Their approximate expressions can be relatively easily found analytically by seeking the $U$-potential minimum under the conditions $r_0,d\!<\!1$ and $\Delta\rho_{\!X^{^{\pm}}}\!>\!1$ consistent with Eq.~(\ref{Drho0Xpm}). This leads to $R_\texttt{min}\!\approx\!(r_{ee}+r_{hh})/2$ and $U_\texttt{min}\!\approx\!-1/d+1/r_{ee}+1/r_{hh}$, where $r_{ee}\!=\!(\lambda/\sigma)\Delta\rho_{\!X^{^{-}}}$ and $r_{hh}\!=\!\lambda\Delta\rho_{\!X^{^{+}}}$ are the inter\-electron and interhole distances in the negative and positive CIE, respectively.~These expressions are seen to reproduce the numerical calculations quite well, within the approximations used, to demonstrate the fast drop of $|U_\texttt{min}|$ (and $T_{cN}$ for the unlike-charge trion crystallization transition, accordingly) with $R_\texttt{min}$ slowly rising as the interlayer separation $d$ in the heterostructure increases.

\subsection{Like-charge trion Wigner crystallization}

In hetero\-structures of two monolayers with $\sigma\!\ne\!1$ separated by an interlayer distance \emph{not} to fulfill the $E_{\!X^{^{-\!}}}(\sigma,d)\!=\!E_{\!X^{^{+\!}}}(\sigma,d)$ constraint, including electrostatically doped heterostructures, either $X^{\!-}$ or $X^{+\!}$ trions are most likely to form under intense irradiation. As can be seen from Fig.~\ref{fig2}, for $\sigma\!<\!1$ the domains $|E_{\!X^{^{-\!}}}|\!>\!|E_{\!X^{^{+\!}}}|$ and $|E_{\!X^{^{-\!}}}|\!<\!|E_{\!X^{^{+\!}}}|$ are located at smaller and greater $d$ to form like-charge trions --- negative and positive, respectively, as long as their binding energy absolute values exceed the thermal fluctuation energy at a given~$T$.

An ensemble of repulsively interacting particles (or quasi\-particles, structureless or compound) forms a~Wig\-ner lattice when its average potential interaction energy exceeds average kinetic energy, $\langle V\rangle/\langle K\rangle\!=\!\Gamma_0\!>\!1$. This was previously shown for systems such as 2D electron gas~\cite{Platzman74}, cold polar molecules~\cite{Buchler07}, and indirect excitons~\cite{LozovikPRL07}. For like-charge trions in Fig.~\ref{fig1}~(b), the Coulomb repulsion at large $R$ is strengthened at shorter $R$ by the dipole-dipole repulsion of their collinear permanent dipole moments (to result in the pairwise interaction potential $V$ illustrated in Fig.~\ref{fig3}), while the total kinetic energy is additionally contributed by the rotational term $K^{\texttt{(r)}}_{\!X^{^{\pm\!}}}\!=\hbar^2l(l+1)/2I_{\!X^{^{\pm}}}$ with $l\!=\!0,1,2,...$ being the orbital quantum number and $I_{\!X^{^{\pm}}}\!\!=\!m_{h,e}r_{hh,ee}/2$ representing the moment of inertia for CIE rotation about its permanent dipole moment direction. The low-$T$ statistical averaging over $l$ leads to the characteristic rotational motion "freezing" temperature $T^{\texttt{(r)}}_{X^{^{\pm\!}}}\!=\!\hbar^2\!/k_BI_{\!X^{^{\pm}}}$~(see, e.g., Ref.~\cite{Pathria}). By direct analogy with the hydrogen molecular ion problem this can be rewritten as $T^{\texttt{(r)}}_{\!X^{^{+\!}}}\!\approx\!\sigma|E_{\!X^{^{+\!}}}|/k_B$ and $T^{\texttt{(r)}}_{\!X^{^{\!-\!}}}\!\approx\!|E_{\!X^{^{\!-\!}}}|/k_B\sigma$ (see, e.g, Ref.~\cite{AbersQM}), indicating the rotational degrees of freedom to be frozen out (at least for the case of $\sigma$ being close to unity typical of TMDs, in particular~\cite{Lambrecht12,Ramasubr12}) as long as the CIEs are stable against the thermal fluctuations.

With no rotational term contribution, it is straightforward to get a qualitative picture of the like-charge trion Wigner crystallization by performing an analysis analogous to that done in Ref.~\cite{Platzman74} for the 2D electron gas. With slight modifications to include the dipole repulsion in the interparticle interaction potential $V$ and to replace the electron mass by the CIE mass in the translational kinetic energy $K$, the expressions for the zero-$T$ critical density $n_{c}$ and for the critical temperature $T^\texttt{(W)}_{c}$ of the Wigner crystallization phase transition take the form (see \emph{Methods})
\begin{eqnarray}
n_{cX^{\pm}}\!=\!\frac{2}{\pi d^2}\!\left(\!\frac{g_{\pm}\Gamma_{0}}{4d}\!\right)^{\!\!2}\!\left[
1-\frac{1}{2}\!\left(\frac{4d}{g_{\pm}\Gamma_{0}}\right)^{\!\!2}\!-\sqrt{1\!-\!\left(\frac{4d}{g_{\pm}\Gamma_{0}}\right)^{\!\!2}}\,\right]\!,\nonumber\\[0.25cm]
k_{B}T^\texttt{(W)}_{cX^{\pm}}=\frac{4Ry^\ast}{g_{\pm}\Gamma_0^2}\,,\hskip3cm\label{ncTc}\\[0.25cm]
g_{\pm}(\sigma)=\left(3+\Big\{\!\begin{array}{c}1\\[-0.25cm]2\end{array}\!\Big\}\,\sigma+\Big\{\!\begin{array}{c}2\\[-0.25cm]1\end{array}\!\Big\}\,
\frac{1}{\sigma}\right)^{\!\!-1}\!\!.\hskip1.6cm\nonumber
\end{eqnarray}
The quantities $n_{cX^{\pm}}$ and $T^\texttt{(W)}_{cX^{\pm}}$ are shown on the top and bottom of Fig.~\ref{fig3}~(c) as functions of $d$ and $\sigma\,(\!<\!1)$, respec\-tively, for moderate $\Gamma_0$ values~\cite{Platzman74}. As $d$ increases so does $\langle V\rangle$ once the dipole repulsion becomes appreciable. With constant $\Gamma_0$ this leads to the $\langle K\rangle$ increase and $n_{cX^{\pm}}$ rise, accordingly. The latter is slightly lower for the negative CIE due to its smaller $K$ because of the smaller mass than that of the positive CIE. Lowering $\sigma$ generally lowers the CIE mass thus decreasing its $K$ whereby $T^\texttt{(W)}_{cX^{\pm}}$ increases. These are the general trends featured in Fig~\ref{fig3}~(c).

\subsection{Estimates for the effects discussed}

We consider the case of the CIE formation in TMD \emph{homo}bilayers (both monolayers of the same material) encapsulated in bulk hexagonal boron nitride (hBN), a popular practical realization one encounters in a wide range of experiments~\cite{Lius-PKim19,Geim20,BondSnoke20,Crooker19}. \emph{Hetero}bilayers (two different TMD monolayers) offer many more CIE formation possibilities and therefore preferably should be analyzed individually. For the quantitative description of the effects predicted, our model requires the knowledge of the exciton reduced effective mass $\mu$, the electron-hole effective masses $m_{e,h}$ associated with it, the \emph{effective} average dielectric constant $\varepsilon$ of the system, and the screening length parameter $r_0\!=\!2\pi\chi_\texttt{2D}$ with $\chi_\texttt{2D}$ being a spatially dispersive (and so nonlocal, i.e. in-plane distance-dependent) polarizability function~\cite{Rubio11,BondMouShal18}. We use $\mu$, $m_e$ and $m_h$ reported recently from the first-principles calculations of the TMD-monolayer electronic structure~\cite{TMDmass18}. The effective dielectric permittivity $\varepsilon$ can be evaluated by the Maxwell-Garnett method~\cite{MG}, which in our case prescribes to use the weighted average of the hBN and TMD static permittivities, whereby for the hBN-monolayer number much greater than two we obtain $\varepsilon\!=\!5.87$ (bulk hBN permittivity averaged over all three directions~\cite{Laturia18}). Finally, the $r_0$ parameter can be obtained based on the fundamental energy minimum principle~\cite{Chandler}, whereby the (negative) binding energy of a CIE complex must contribute \emph{the most} in order for the CIE ensemble to be at a local minimum of its total energy in equilibrium. The $r_0$ parameter can therefore be found as the maximum point of the CIE binding energy absolute value $|E_{\!X^{^{\pm}}}(\sigma,d,r_0)|$ taken with both $\sigma\!=\!m_e/m_h$ and $d$ fixed. We note that by its definition the KR potential screening length $r_0$ refers to in-plane charges which are the like-charge carriers to form the CIE in our case. These carriers are separated by distances at least of the order of $2a_B^{\ast}$ --- much greater than those of the order of $a_B^{\ast}$ one typically encounters in the exciton case. Therefore, being determined by greater distances, our $r_0$ due to its inherent nonlocality may very well be different from the values previously reported theoretically and experimentally for excitons in TMD monolayers~\cite{Berkelbach2013,Crooker19}.

\begin{figure}[t]
\includegraphics[scale=0.82]{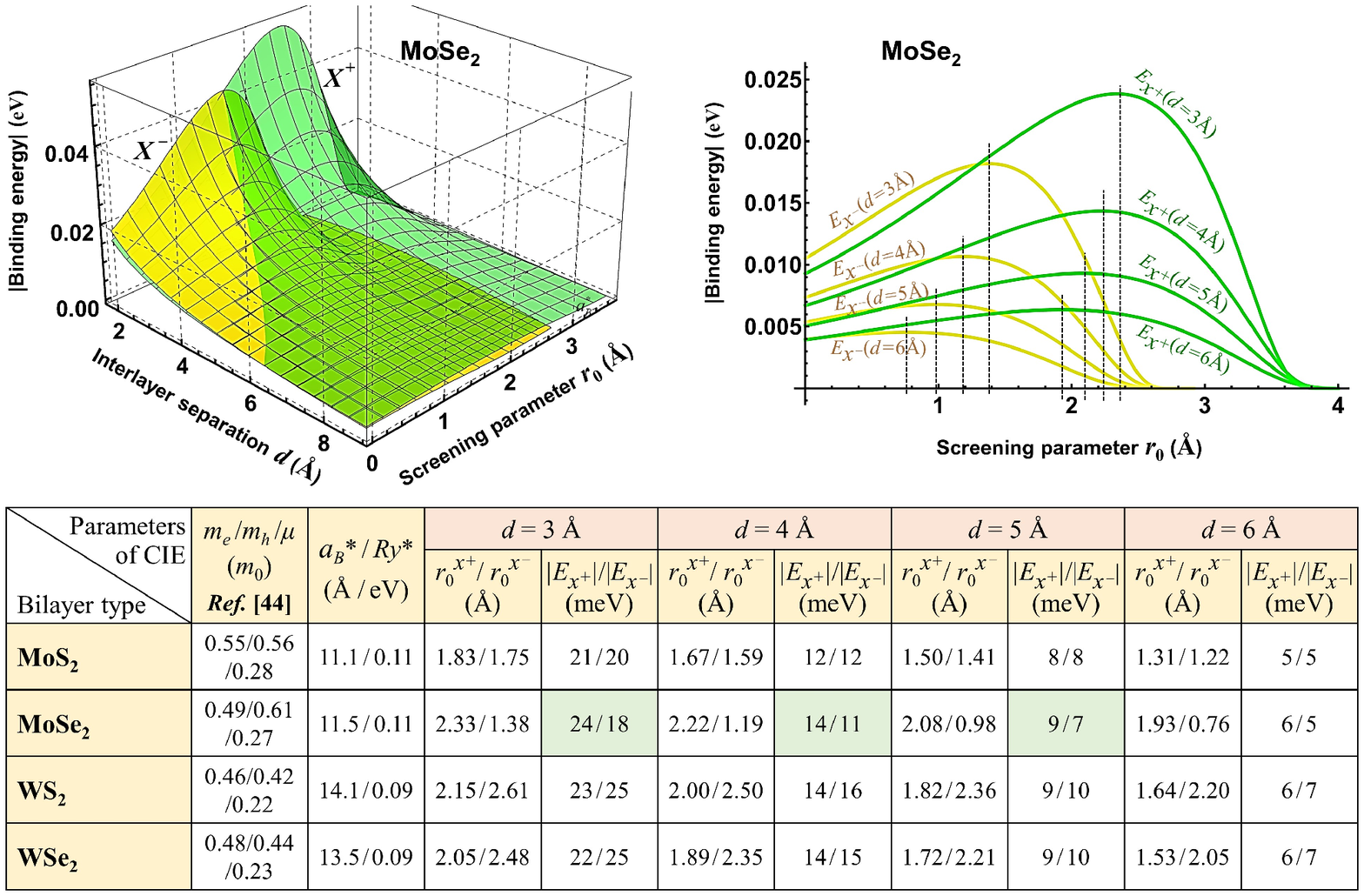}
\caption{An example of the MoSe$_2$ homobilayer embedded in bulk hBN material (\emph{top, left and right panels}) to present the positive ($X^{+}$) and negative ($X^{-}$) CIE binding energy surfaces as functions of the interlayer distance $d$ and the screening length parameter $r_0$ (\emph{left panel}), and their fixed-$d$ crosscuts as functions of $r_0$ (\emph{right panel}). The vertical dashed lines on the top right panel indicate the maxima points to give the actual $X^{+}$ and $X^{-}$ binding energy absolute values at the interlayer distances $d$ fixed. The CIE parameters thus obtained are tabulated at the bottom for the four types of homobilayers chosen. Highlighted greenish in the table are the largest differences between the $X^{+}$ and $X^{-}$ binding energy absolute values.}
\label{FigTable}
\end{figure}

Figure~\ref{FigTable} summarizes the data we have obtained for the hBN-encapsulated homobilayers of MoS$_2$, MoSe$_2$, WS$_2$ and WSe$_2$. With $\mu$, $m_{e,h}$ and $\varepsilon$ found as described above, we first calculate the exciton Bohr radius $a_B^\ast$ and Rydberg energy $Ry^\ast$ for each case individually. Then, with known $\sigma\!=\!m_e/m_h$, $a_B^\ast$ and $Ry^\ast$ we obtain the binding energy surfaces $|E_{\!X^{^{\pm}}}(\sigma,d,r_0)|$ in physical units from Eqs.~(\ref{bindingE}),~(\ref{JXfin}) and (\ref{Drho0Xpm}), determine their maximum points $r_0^{X^{\pm}}$ for a particular fixed $d$, and compute the CIE binding energy absolute values $|E_{\!X^{^{\pm}}}(\sigma,d,r_0^{X^{\pm}})|$. We do this for the interlayer distances $d\!=\!3,4,5$~and~$6\,$\AA$\,$ (typical of van der Waals coupling) for each homobilayer type in order to be able to see the tendencies for the $X^+$ and $X^-$ trion formation as $d$ increases. As an example, the left and right top panels in Fig.~\ref{FigTable} show the $X^+$ and $X^-$ binding energy surfaces and their fixed-$d$ crosscuts, respectively, for the MoSe$_2$ homobilayer. The vertical dashed lines on the right panel trace the $|E_{\!X^{^{\pm}}}|$ maxima and their respective $r_0^{X^{\pm}}$ distances. The CIE parameters thus obtained are tabulated at the bottom of Fig.~\ref{FigTable} for all four homobilayers selected. We note the general consistency of our $|E_{\!X^{^{\pm}}}|$ obtained both with numerical simulation data reported previously for the MoS$_2$/WS$_2$ heterobilayer embedded in hBN ($18/28$~meV for the $X^+/X^-$ trion~\cite{Thygesen18}) and with the latest experimental observations on the MoSe$_2$/WSe$_2$ hetero\-bilayer system ($10/15$~meV for the $X^+/X^-$ trion~\cite{Lius-PKim19} and $28$~meV for the $X^-$ trion~\cite{Geim20}, respectively). Highlighted greenish in the table are the largest differences between the positive and negative trion binding energies in MoSe$_2$ due to a significant $m_e$ and $m_h$ difference yielding $\sigma\!=0.8$, which makes this homobilayer energetically favorable for the positive CIE Wigner crystallization for the interlayer distances $d$ ranging between $3$ and $5$~\AA. As $d$ increases from $3$ to $6$~\AA, for all types of bilayers tabulated, both $|E_{\!X^+}|$ and $|E_{\!X^-}|$ quickly decrease and get closer together while still remaining significant in magnitude, to make the normal unlike-charge trion crystallization energetically favorable. In the case of MoSe$_2$, this implies a crossover from the Wigner crystal phase of the \emph{positive} trions to the normal crystal phase of the unlike-charge trions. A~similar crossover from the Wigner crystallization of the \emph{negative} trions to the normal crystallization of the unlike-charge trions, although not as pronounced as for MoSe$_2$, might also be the case for WS$_2$ and WSe$_2$ according to our data tabulated. For MoS$_2$, on the contrary, only the normal unlike-charge trion crystallization is energetically favorable as $|E_{\!X^+}|$ and $|E_{\!X^-}|$ there are about the same over the entire range of the interlayer distances $d$ presented.

Using $a_B^\ast$ and $Ry^\ast$ obtained as scaling units, it is quite straightforward to estimate the critical parameters for many-particle CIE systems in TMD homobilayers tabulated in Fig~\ref{FigTable}. As our scaling units are very close for all homobilayers presented (an immediate corollary of being embedded in bulk hBN), from Fig.~\ref{fig3}~(b) and (c) one can get $k_BT^\texttt{(N)}_{c}\!\approx|U_{\texttt{min}}|\!\approx0.3\;$eV at $d\!=6\;$\AA, critical density $n_{cX^{\pm}}\!\approx\!10^{12}\!\div10^{13}\,$cm$^{-2}$ and $k_BT^\texttt{(W)}_{cX^{\pm}}\!\approx6\;$meV (to give $T^\texttt{(W)}_{cX^{\pm}}\!\approx\!70\;$K). The fact of $k_BT^\texttt{(N)}_{c}$ being much greater than our $|E_{\!X^{^{\pm\!}}}|$ tabulated tells that the dipole-ordered normal 1D-crystal phase is the actual ground state of the many-particle unlike-charge trion system. The obtained $n_{cX^{\pm}}$ and $T^\texttt{(W)}_{cX^{\pm}}$ are, respectively, close to and exceed those reported experimentally for IEs~\cite{Li17,Lius-PKim19}, suggesting that the Wigner crystallized CIE phase can be realized in properly fabricated vdW heterostructures with the twofold overbalance of negative [as in Fig.~\ref{fig1}~(b)] or positive charge carriers. Crystallized exciton photoemission features can be found in Ref.~\cite{Suris16}.

\emph{In summary}, we study the properties of charged interlayer excitons in highly excited vdW heterostructures --- a compound fermion system with the permanent dipole moment observed recently in TMD bilayers~\cite{Lius-PKim19,JonFinley}. We predict the existence of new strongly correlated collective CIE states, the long-range ordered phases of the excited heterostructure --- the crystal phase and the Wigner crystal phase. We evaluate the critical temperatures and density for the formation of such many-particle cooperative compound fermion states. We demonstrate that they can be selectively realized with bilayers of properly chosen electron-hole effective mass ratio by just varying their interlayer separation distance. Compound fermion systems featuring permanent electric dipole moments are of both fundamental and practical importance due to their inherently unique many-body correlation effects between electric-dipole and spin degrees of freedom. The spin in such systems could potentially be used for quantum information processing and its correlation with the dipole moment provides an opportunity for spin manipulation through optical means. Fundamental cooperative crystallization phenomena we predict herewith will greatly increase the potential capabilities of such systems to open up new avenues for experimental exploration and novel device technologies with vdW hetero\-structures.

\section{Methods}

\subsection{The charged interlayer exciton binding energy}

\begin{figure}[b]
\includegraphics[scale=0.43]{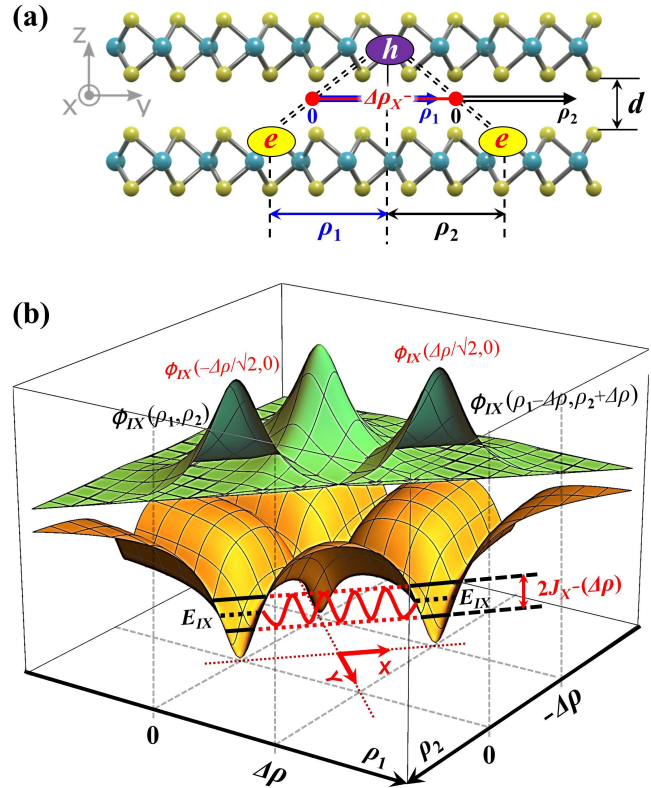}\vskip-0.65cm
\caption{(a)~The structure of a negatively charged interlayer exciton (trion) in a TMD bilayer. (b)~Schematic (a diagonal crosscut) of the tunnel exchange coupling configuration to form the interlayer trion complex sketched in~(a). The coupling occurs in the configuration space of the two \emph{independent} in-plane relative $e$-$h$ motion coordinates $\rho_1$ and $\rho_2$ representing the excitons separated by the center-of-mass-to-center-of-mass distance $\Delta\rho$ ($=\!\Delta\rho_{X^{^{-}}}$ for the negative CIE). The coupling is due to the tunneling of the system through the potential barrier formed by the two single-exciton Coulomb interaction potentials (bottom, yellow) given by the second line in Eq.~(\ref{Ham}), between the equivalent states (top, green) represented by the isolated two-exciton wave functions in Eq.~(\ref{phixy}).}
\label{fig4}
\end{figure}

A sketch of a charged interlayer exciton (CIE, or trion) in a TMD bilayer is presented in Fig.~\ref{fig4}~(a) for the negative trion case ($X^-$). The positive trion case ($X^+$) can be obtained by the charge sign inversion. The CIE we deal with here is a charged three-particle complex of an interlayer (indirect) exciton (IE) and an extra hole ($h$) or electron ($e$), in which two like charge carriers confined to the same layer share an unlike charge carrier on the other layer. Such a CIE complex can be viewed as being formed by the \emph{two} equivalent indistinguishable symmetric IE configurations with an extra charge carrier attached to the left or right IE, respectively, as shown in Fig.~\ref{fig4}~(a) for the negative trion case~\cite{Bondarev2016}. For such a quantum system the effective configuration space can be represented by the two \emph{independent} in-plane projections $\rho_1$ and $\rho_2$ of the relative $e$-$h$ coordinates (relative to the center of mass) of each of the IEs, whereby the $X^\pm$ ground-state Hamiltonian takes the following form~\cite{BondVlad18}
\begin{eqnarray}
\hat{H}(\rho_1,\rho_2,\Delta\rho,d)=-\frac{1}{\rho_1}\frac{\partial}{\partial\,\!\rho_{1}}\,\rho_1\frac{\partial}{\partial\,\!\rho_{1}}
-\frac{1}{\rho_2}\frac{\partial}{\partial\,\!\rho_{2}}\,\rho_2\frac{\partial}{\partial\,\!\rho_{2}}\hskip3.5cm\label{Ham}\\[0.2cm]
+\,V_\texttt{C}\Big(\!\sqrt{\rho_1^2+d^2}\,\Big)+\,V_\texttt{C}\Big(\!\sqrt{\rho_2^2+d^2}\,\Big)
+\,V_\texttt{C}\Big(\!\!\sqrt{(\rho_1\pm\Delta\rho)^2+d^2}\Big)+\,V_\texttt{C}\Big(\!\!\sqrt{(\rho_2\mp\Delta\rho)^2+d^2}\Big)\nonumber\\[0.2cm]
+\,2\left\{\!\!\begin{array}{lcr}V_\texttt{KR}(|\sigma(\rho_1-\rho_2)/\lambda+\Delta\rho|) & \;\longrightarrow\; & X^+\\
V_\texttt{KR}(|(\rho_1-\rho_2)/\lambda-\Delta\rho|) & \;\longrightarrow\; & X^-\end{array}\right..\hskip4.3cm\nonumber
\end{eqnarray}
The "atomic units" are used with distance and energy measured in the units of exciton Bohr radius $a^\ast_B\!=0.529\,\mbox{\AA}\,\varepsilon/\mu$ and Rydberg energy $Ry^\ast\!\!=\!\hbar^2\!/(2\mu\,m_0a_B^{\ast2})\!=\!e^2\!/(2\varepsilon a_B^\ast)\!=\!13.6\,\mbox{eV}\,\mu/\varepsilon^2$, respectively~\cite{LandauQM,Pitaevski63,Herring,LeavittLittle}, $\mu\!=\!m_e/(\lambda\,m_0)$ with $\lambda\!=\!1+\sigma$ stands for the exciton reduced effective mass (in the units of free electron mass $m_0$), $\sigma\!=\!m_e/m_h$ is the electron-to-hole effective mass ratio, and $\varepsilon$ represents the \emph{effective} average dielectric constant of the entire bilayer structure~\cite{LeavittLittle}. The image-charge effects are neglected.

The first two lines in Eq.~(\ref{Ham}) describe the kinetic and potential energy, respectively, for the two non-interacting IEs. Their individual $e$-$h$ attractive Coulomb potentials \emph{screened}, generically of the form
\begin{equation}
V_\texttt{C}(r)=-\frac{1}{r}=-\frac{1}{\sqrt{\rho^2+d^2}}
\label{VC}
\end{equation}
(atomic units) with $\rho$ being the in-plane intercharge distance, are symmetrized to account for the presence of the neighbor a distance~$\Delta\rho$ away as seen from the $\rho_1$- and $\rho_2$-coordinate systems assigned to originate at the respective IE centers-of-mass and treated independently; see Fig.~\ref{fig4}~(a). The last line is the interexciton exchange Coulomb interaction (or the like-charge Coulomb repulsion potential inside the trion) --- $h$-$h$ for $X^+$ and $e$-$e$ for $X^-$, respectively. We use the repulsive KR interaction potential to represent this interaction (atomic units)
\begin{equation}
V_\texttt{KR}(\rho)=\frac{\pi}{(\epsilon_1+\epsilon_2)r_0}\left[H_{0}\!\left(\frac{\rho}{r_0}\right)-N_{0}\!\left(\frac{\rho}{r_0}\right)\right],
\label{VKR}
\end{equation}
in order to properly take into account the screening effect for the like charges confined to the same monolayer~\cite{KeldyshRytova}. Here, $N_0$ and $H_0$ are the $0$th order Neumann and Struve functions, respectively, $r_0$ is the screening length defined in Eq.~(\ref{keldyshpot}) for a 2D material~\cite{Berkelbach2013}, and $\epsilon_{1,2}$ are the dielectric permittivities of its surroundings. To facilitate the analytical calculations, we approximate Eq.~(\ref{VKR}) by its accurate alternative (\ref{keldyshpot}) written in terms of elementary functions as discussed and proposed for atomically thin layers in Ref.~\cite{Rubio11}.

For the CIE complex of two identical configurations with an extra charge attached to the left or right IE, the total wave function must be either symmetric or antisymmetric with respect to their interchange due to the conservation of parity. This can generally be achieved with coordinate wave functions of the form
\begin{equation}
\Psi_{g,u}\sim\frac{1}{\sqrt{2}}\left[\phi_{I\!X}(\rho_1,\rho_2)\pm\phi_{I\!X}(\rho_2,\rho_1)\right]=\frac{1}{\sqrt{2}}\left[\phi_{I\!X}(\rho_1,\rho_2)\pm\phi_{I\!X}(\rho_1-\Delta\rho,\rho_2+\Delta\rho)\right],
\label{IXwfuncpm}
\end{equation}
where $\phi_{I\!X}(\rho_1,\rho_2)=\psi_{I\!X}(\rho_1,d)\,\psi_{I\!X}(\rho_2,d)$ with $\psi_{I\!X}$ being the IE wave function. This involves the two terms localized at $\rho_1\!=\!\rho_2\!=\!0$ and $\rho_1\!=\!-\rho_2\!=\!\Delta\rho$, respectively, to represent the two equivalent configurations in terms of the two independent relative $e$-$h$ coordinates $\rho_1$ and $\rho_2$ as shown in Fig.~\ref{fig4}~(a). Since the total wave function of the quantum ground state must be nodeless~\cite{LandauQM}, for large $\Delta\rho\gg1$ the ground-state wave function of two IEs (two bosons) must be symmetric in coordinates to hold with $\Psi_g$ in Eq.~(\ref{IXwfuncpm}). At shorter $\Delta\rho\gtrsim1$ it can be multiplied by an \emph{even} function of coordinates to be found from the Hamiltonian~(\ref{Ham}) in the manner similar to that developed in the past for the hydrogen molecule and molecular ion in seminal works by Landau, Gor'kov, Pitaevski, Holstein and Herring~\cite{LandauQM,Pitaevski63,Herring} and more recently by one of us for biexcitons and trions in quasi-1D/2D semiconductors~\cite{BondVlad18,Bondarev11PRB,Bondarev2016}. Assuming further that for both configurations their respective IEs are in the spin-singlet states as dictated by the hyperfine interactions of their unlike-charge spin-1/2 fermionic (electron and hole) constituents~\cite{AbersQM}, one arrives at the CIE complex featuring the ground state with two identical like-charge collinear-spin fermions in the same layer, which are thereby forced both by the Coulomb repulsion and by the Pauli exclusion principle to avoid each other at short $\Delta\rho<1$. Such a CIE complex is therefore only possible to form due to the asymptotic Coulomb exchange coupling at $\Delta\rho\gtrsim1$, the domain our theory applies for.

Figure~\ref{fig4}~(b) shows a diagonal vertical crosscut of the potential energy surface (bottom) as given for $X^-$ by the second line of Eq.~(\ref{Ham}) in the two-coordinate configuration space $(\rho_1,\rho_2)$. On the main diagonal, this surface has two symmetrical minima separated by the potential barrier. The minima represent the two \emph{equivalent} isolated IE states (top) given by the solution to the ground-state eigenvalue problem defined by the first two lines of the Hamiltonian~(\ref{Ham}). This solution is the product of the two ground-state IE wave functions. The interlayer (or indirect) exciton eigenvalue problem was previously studied by Leavitt and Little~\cite{LeavittLittle}. Their ground-state energy $E_{I\!X}$ and the wave-function $\psi_{I\!X}$ are as follows (atomic units)
\begin{equation}
E_{I\!X}(d)=\alpha^2-\frac{4\alpha+4\alpha^4d^2E_1(2\alpha d)\exp(2\alpha d)}{1+2\alpha d}\,,
\label{IXEn}
\end{equation}
where $E_1(x)\!=\!\int_{x}^{\infty}\!dt\,e^{-t}\!/t$ is the exponential integral, $\alpha\!=\!2/(1+2\sqrt{d}\,)$, and
\begin{equation}
\psi_{I\!X}(\rho,d)=N\exp[-\alpha(\sqrt{\rho^2+d^2}-d)]\,,
\label{IXwfunc}
\end{equation}
with $N\!=4/\sqrt{1+4\sqrt{d}+8d\,(1+\sqrt{d}\,)}\,$ as per the normalization $\int_{0}^{\infty}\!d\rho\,\rho\,|\psi_{I\!X}(\rho,d)|^2=1$.

As described at large in Refs.~\cite{Bondarev2016,BondVlad18}, we start the CIE binding energy calculation with the $(\rho_1,\rho_2)$-configuration space transformation to the new coordinates as follows
\begin{equation}
x=\left\{\!\!\begin{array}{lcr}(\rho_2-\rho_1-\Delta\rho)/\sqrt{2} & \;\longrightarrow\; & X^+\\
(\rho_1-\rho_2-\Delta\rho)/\sqrt{2} & \;\longrightarrow\; & X^-\end{array}\right.,\hskip1.cm y=\frac{\rho_1+\rho_2}{\sqrt{2}}\,.
\label{transformation}
\end{equation}
This transformation places the origin and both axes of the new coordinate system $(x,y)$ as shown in Fig.~\ref{fig4}~(b) --- in the middle of the potential barrier that separates the two potential wells representing the two equivalent isolated IE states --- to capture the maximal tunnel flow $J_{X^{^\pm}}(\Delta\rho)$ between the two indistinguishable IE configurations. An approximate solution to the Schr\"{o}dinger equation with the Hamiltonian~(\ref{Ham}) can be constructed using Eq.~(\ref{IXwfunc}). By converting Eq.~(\ref{IXwfunc}) to the $(x,y)$-space per Eq.~(\ref{transformation}), we define the product wave function
\begin{eqnarray}
\phi_{I\!X}(x,y)=\psi_{I\!X}[\rho_1(x,y),d]\,\psi_{I\!X}[\rho_2(x,y),d],\hskip0.5cm\nonumber\\[-0.25cm]
\label{phixy}\\[-0.25cm]
\rho_{1,2}(x,y)=\left\{\!\!\begin{array}{lcr}(y\mp x)/\sqrt{2}\mp\Delta\rho/2 & \;\longrightarrow\; & X^+\\
(y\pm x)/\sqrt{2}\pm\Delta\rho/2 & \;\longrightarrow\; & X^-\end{array}\right.,\nonumber
\end{eqnarray}
to describe the motion with the energy $E_{I\!X}$ inside the potential well centered at $\rho_1\!=\!\rho_2\!=\!0$ (or $x\!=\!-\Delta\rho/\sqrt{2}$, $y\!=\!0$), while being exponentially damped outside. In just the same way, the function $\phi_{I\!X}(-x,y)$ describes the motion with the same energy inside the well centered at $\rho_1\!=-\!\rho_2\!=\!\Delta\rho$ for the $X^-$ case shown in Fig.~\ref{fig4}~(b) and at $\rho_2\!=-\!\rho_1\!=\!\Delta\rho$ for the $X^+$ case (both corresponding to $x\!=\!\Delta\rho/\sqrt{2}$, $y\!=\!0$). Both of these functions are properly normalized to unity within their respective potential wells. Both of them are even in $x$ and $y$ with respect to their respective well-center positions, whereby $\partial \phi_{I\!X}(\mp\Delta\rho/\sqrt{2},y)/\partial x\!=\partial \phi_{I\!X}(x,0)/\partial y\!=0$.

When the small probability of the underbarrier tunneling is taken into account, the energy level $E_{I\!X}$ splits into $E_{I\!X}-J_{X^{^\pm}}(\Delta\rho)$ and $E_{I\!X}+J_{X^{^\pm}}(\Delta\rho)$. Then, the correct \emph{zero}-approximation wave functions corresponding to these levels are $\left[\phi_{I\!X}(x,y)\pm\phi_{I\!X}(-x,y)\right]/\sqrt{2}$, and since $\phi_{I\!X}(x,y)\phi_{I\!X}(-x,y)$ is vanishingly small everywhere, they are normalized so that the integrals of their squares over \emph{both} wells are unity. This suggests that the actual eigenfunctions of the eigenvalues $E_{g,u}$ can be written as
\begin{equation}
\psi_{g,u}(x,y)=\frac{1}{\sqrt{2}}\left[\psi_{X^{^\pm}}(x,y)\pm\psi_{X^{^\pm}}(-x,y)\right],
\label{psigu}
\end{equation}
where $\psi_{X^{^\pm}}(-\Delta\rho/\sqrt{2},y)\!=\!\phi_{I\!X}(-\Delta\rho/\sqrt{2},y)$, with the unknown function $\psi_{X^{^\pm}}(x,y)$ representing an approximate solution to the Schr\"{o}dinger equation with the Hamiltonian (\ref{Ham}) brought to the $(x,y)$-space per Eq.~(\ref{transformation}) to take the form
\begin{equation}
\hat{H}(x,y,\Delta\rho,d)=\hat{T}(x,y,\Delta\rho)+U(x,y,\Delta\rho,d).
\label{Hamxy}
\end{equation}
Here the kinetic and potential energy terms are as follows
\begin{eqnarray}
\hat{T}=-\frac{\partial^2}{\partial x^2}-\frac{\partial^2}{\partial y^2}-2\frac{(x\!+\!\Delta\rho/\sqrt{2})\,\partial\!/\partial x-y\,\partial\!/\partial y}{(x\!+\!\Delta\rho/\sqrt{2})^2-y^2}\,,\hskip3.5cm\nonumber\\
\label{Txy}\\
U=\!\sum_{\alpha,\beta=0}^1\!V_\texttt{C}\Big\{\!\sqrt{\big[x\!+\!(-1)^\alpha\Delta\rho/\sqrt{2}\!+\!(-1)^\beta y\big]^2\!/2\!+d^2}\,\Big\}
+\,2\left\{\!\!\begin{array}{lcr}V_\texttt{KR}\big(|\sqrt{2}\sigma x-\Delta\rho|/\lambda\big) & \longrightarrow & X^+\\
V_\texttt{KR}\big(|\sqrt{2}x-\sigma\Delta\rho|/\lambda\big) & \longrightarrow & X^-\end{array}\right..\nonumber
\end{eqnarray}

In general, the function $\psi_{X^{^\pm}}(x,y)$ is supposed to preserve the parity and the behavior of the function $\phi_{I\!X}(x,y)$, to \emph{only} depart noticeably from $\phi_{I\!X}(x,y)$ in the \emph{very} tail area $x\!\sim\!y\!\sim\!0$ under the potential barrier and to overlap with $\psi_{X^{^{\pm}}}(-x,y)$ in there; see Fig.~\ref{fig4}~(b). The overlap enables the tunnel exchange between the two indistinguishable configurations represented by $\phi_{I\!X}(x,y)$ pinned to the potential well centered at $\rho_1\!=\!\rho_2\!=\!0$ ($x\!=\!-\Delta\rho/\sqrt{2}$, $y\!=\!0$) and by $\phi_{I\!X}(-x,y)$ pinned to the other potential well at $\rho_1\!=\!-\rho_2\!=\!\Delta\rho$ or $\rho_2\!=\!-\rho_1\!=\!\Delta\rho$ ($x\!=\!\Delta\rho/\sqrt{2}$, $y\!=\!0$) for $X^-$ and $X^+$, respectively. Under these restrictive assumptions about $\psi_{X^{^\pm}}(x,y)$ in Eq.~(\ref{psigu}), it is possible to write down the two Schr\"{o}dinger equations as follows
\[
(\hat{T}\!+U)\psi_{X^{^\pm}}(x,y)=2E_{I\!X}\psi_{X^{^\pm}}(x,y),\hskip0.5cm(\hat{T}+U)\psi_g(x,y)=E_g\psi_g(x,y),
\]
where $\hat{T}$ and $U$ are those of Eq.~(\ref{Txy}). We multiply from the left the former by $\psi_g(x,y)$ and the latter by $\psi_{X^{^\pm}}(x,y)$, subtract one from another, and integrate over $x$ from $-\infty$ to $0$ and over $y$ from $-\infty$ to $+\infty$. This includes the potential well positioned at $x\!=\!-\Delta\rho/\sqrt{2}$, $y\!=\!0$, so that
\[
\int_{\!-\infty}^{0}\!\!\!\!\!\!dx\!\int_{\!-\infty}^{\infty}\!\!\!\!\!\!dy\,\psi_{X^{^\pm}}(x,y)\psi_g(x,y)=\!
\frac{1}{\sqrt{2}}\int_{\!-\infty}^{0}\!\!\!\!\!\!dx\!\int_{\!-\infty}^{\infty}\!\!\!\!\!\!dy\,\psi_{X^{^\pm}}^2(x,y)\approx\!
\frac{1}{\sqrt{2}}\int_{\!-\infty}^{0}\!\!\!\!\!\!dx\!\int_{\!-\infty}^{\infty}\!\!\!\!\!\!dy\,\phi_{I\!X}^2(x,y)=\!\frac{1}{\sqrt{2}}\,,
\]
and we find
\[
2E_{I\!X}-E_g=\sqrt{2}\int_{\!-\infty}^{0}\!\!\!\!\!dx\!\int_{\!-\infty}^{\infty}\!\!\!\!\!dy\Big[\psi_g(x,y)\hat{T}\psi_{X^{^\pm}}(x,y)
-\psi_{X^{^\pm}}(x,y)\hat{T}\psi_g(x,y)\Big].
\]
In here, with $T$ of Eq.~(\ref{Txy}) it can be seen that its last term might only be significant at or close to $x\!=\!-\Delta\rho/\sqrt{2}$, $y\!=\!0$, but the partial derivatives of relevance are zero there, and so this term can be dropped for smallness over the entire integration domain. What remains can be integrated by parts. Bearing in mind that $\psi_g(0,y)\!=\!\sqrt{2}\,\psi_{X^{^\pm}}(0,y)$, $\partial \psi_g(0,y)/\partial x\!=\!0$ and all the functions involved as well as their derivatives must vanish at infinity, this after numerus cancelations gives
\[
2E_{I\!X}-E_g=2\!\int_{\!-\infty}^{\infty}\!\!\!\!\!dy\,\psi_{X^{^\pm}}(0,y)\frac{\partial \psi_{X^{^\pm}}(0,y)}{\partial x}\,.
\]
From here, with just a tiny adjustment for practical application purposes, we obtain the tunnel exchange splitting integral in Eq.~(\ref{bindingE}) of the following final form
\begin{equation}
J_{X^{^\pm}}(\Delta\rho)=\int_{\!-\Delta\rho/\!\sqrt{2}}^{\Delta\rho/\!\sqrt{2}}\!dy\left|\psi_{X^{^\pm}}(x,y)\frac{\partial\psi_{X^{^\pm}}(x,y)}{\partial x}\right|_{x=0}.
\label{JXpm}
\end{equation}
Here, we take into account the fast exponential drop-off of the integrand away from the $y\!=\!0$-plane, whereby the integration limits can be shrunken to only include the physically significant cross-section region, see Fig.~\ref{fig4}~(b), that controls the under-barrier tunnel probability flow --- a positive quantity we wish to stress by taking the absolute value of. Such a tunnel exchange coupling binds the three-particle system to form a stable CIE state.

\vskip0.25cm\emph{(a)~The Trion Wave Function}

We seek the function $\psi_{X^{^\pm}}(x,y)$ of Eq.~(\ref{JXpm}) in the following form
\begin{equation}
\psi_{X^{^\pm}}(x,y)=\phi_{I\!X}(x,y)\exp[-S_{X^{^\pm}}(x,y)]\,.
\label{psixy}
\end{equation}
Here, the unknown function $S_{X^{^\pm}}(x,y)$ is to be chosen so that $S_{X^{^\pm}}(x\!=\!-\Delta\rho/\sqrt{2},y)\!=0$ to fulfill the condition $\psi_{X^{^\pm}}(-\Delta\rho/\sqrt{2},y)\!=\!\phi_{I\!X}(-\Delta\rho/\sqrt{2},y)$ as per Eq.~(\ref{psigu}), while also being smooth and \emph{slowly} varying in the domain $|x|,|y|\!<\!\Delta\rho/\!\sqrt{2}$ under the barrier, whereby its second derivatives should be negligible. Additionally, as was mentioned above, for our three-particle $X^\pm$ complexes the equivalency of the two IEs sharing the same hole (or electron) implies their identity and leads to the fact of the like-charge carriers having collinear spins. The Coulomb repulsion strengthened by the Pauli exclusion principle forces them to avoid each other at short interexciton center-of-mass-to-center-of-mass distance $\Delta\rho\!<\!1$, making it possible for a stable CIE complex to only form at $\Delta\rho\!\gtrsim\!1$, which is why $1/\Delta\rho$ can be used as a smallness parameter in analytical calculations.

For the negative CIE, plugging Eq.~(\ref{psixy}) into the Schr\"{o}dinger equation with the Hamiltonian~(\ref{Hamxy}),(\ref{Txy}), to the first non-vanishing order in $1/\Delta\rho$ one obtains
\begin{equation}
\frac{\partial S_{X^{^-}}}{\partial x}\approx\frac{\Delta\rho}{\sqrt{2}\left(\alpha\Delta\rho-1\right)}\,
V_\texttt{KR}\Big(\frac{|\sqrt{2}\,x-\sigma\Delta\rho|}{\lambda}\Big),
\label{SPDEapp1}
\end{equation}
where the second-order derivatives of $S_{X^{^-}}$ are neglected. To find the analytical solution to this differential equation in the domain of interest $|x|,|y|\!<\!\Delta\rho/\!\sqrt{2}$, we use $V_{\texttt{eff}}$ of Eq.~(\ref{keldyshpot}) to replce $V_\texttt{KR}$ in the right-hand side of Eq.~(\ref{SPDEapp1}). The solution to fulfill the boundary condition $S_{X^{^-}}(-\Delta\rho/\sqrt{2},y)\!=\!0$ is then given by
\begin{equation}
S_{X^{^-}}(x,y)=\frac{\Delta\rho}{\sqrt{2}\left(\alpha\Delta\rho-1\right)}
\int_{\!-\Delta\rho/\!\sqrt{2}}^x\!\!\!dt\,V_{\texttt{eff}}\!\Big(\frac{|\sqrt{2}\,t-\sigma\Delta\rho|}{\lambda}\Big)=
\frac{\Delta\rho}{\sqrt{2}\left(\alpha\Delta\rho-1\right)}\,I(x,\Delta\rho).
\label{SXmintegral}
\end{equation}
To calculate the integral $I(x,\Delta\rho)$ here, we first use the unit step function to write
\[
V_{\texttt{eff}}\!\Big(\frac{|\sqrt{2}\,t-\sigma\Delta\rho|}{\lambda}\Big)\!=\theta\Big(t-\frac{\sigma\Delta\rho}{\sqrt{2}}\Big)
V_{\texttt{eff}}\!\Big(\frac{\sqrt{2}\,t-\sigma\Delta\rho}{\lambda}\Big)\!+\theta\Big(\frac{\sigma\Delta\rho}{\sqrt{2}}-t\Big)
V_{\texttt{eff}}\!\Big(\frac{\sigma\Delta\rho-\sqrt{2}\,t}{\lambda}\Big),
\]
followed by the change of variable $\tau\!=\!(\!\sqrt{2}\,t-\sigma\Delta\rho)/\lambda$ to obtain
\begin{eqnarray}
I(x,\Delta\rho)=\frac{\lambda}{\sqrt{2}}\Big\{\theta\Big(x-\frac{\sigma\Delta\rho}{\sqrt{2}}\Big)\Big[
\int_{\!-\Delta\rho}^{0}\!\!\!d\tau\,V_{\texttt{eff}}(-\tau)+\!\int_{0}^{(\sqrt{2}\,x-\sigma\Delta\rho)/\lambda}\hskip-1.5cm d\tau\,V_{\texttt{eff}}(\tau)\Big]\nonumber\\[0.25cm]
+\;\theta\Big(\frac{\sigma\Delta\rho}{\sqrt{2}}-x\Big)\!\int_{\!-\Delta\rho}^{(\sqrt{2}\,x-\sigma\Delta\rho)/\lambda}\hskip-1.5cm d\tau\,V_{\texttt{eff}}(-\tau)
\Big\}\,.\hskip3cm\nonumber
\end{eqnarray}
Of three terms here, only the third is seen to provide the solution in the domain $x\!<\!\sigma\Delta\rho/\!\sqrt{2}$ that includes the region $x\!\sim\!0$ of interest to us. With $V_{\texttt{eff}}$ of Eq.~(\ref{keldyshpot}), this term can be easily calculated analytically using integration by parts. One obtains
\[
I(x\!<\!\sigma\Delta\rho/\!\sqrt{2},\Delta\rho)=\frac{\lambda}{\sqrt{2}}\,\Big[\ln\!\frac{1+p}{1+s}+\ln\!\Big(\frac{s}{1+s}\Big)^{\!s}\!\Big(\frac{1+p}{p}\Big)^{\!p}
+(\ln2-\gamma)(e^{-s}-e^{-p})\Big]
\]
with $s\!=\!(\sigma\Delta\rho-\!\sqrt{2}\,x)/\lambda r_0$ and $p=\!\Delta\rho/r_0$. A close inspection of this expression reveals that since $s\!<\!p\,$, the first summand is predominant there and the other two are negligible for all $1\!<\!s\!<\!p$ regardless of how big $s$ and $p$ individually are. After dropping the negligible terms, Eq.~(\ref{SXmintegral}) in the domain of interest takes the final form as follows
\begin{equation}
S_{X^{^-}}(x,y)\approx\frac{\lambda\Delta\rho}{2\left(1-\alpha\Delta\rho\right)}\,
\ln\!\frac{1-\sqrt{2}\,x/(\lambda r_0+\sigma\Delta\rho)}{1+\Delta\rho/(\lambda r_0+\sigma\Delta\rho)}\,.
\label{SXm}
\end{equation}

For the positive CIE, plugging Eq.~(\ref{psixy}) into the Schr\"{o}dinger equation with the Hamiltonian~(\ref{Hamxy}),(\ref{Txy}) yields to the first non-vanishing order in $1/\Delta\rho$ the equation as follows
\begin{equation}
\frac{\partial S_{X^{^+}}}{\partial x}\approx\frac{\Delta\rho}{\sqrt{2}\left(\alpha\Delta\rho-1\right)}\,
V_\texttt{KR}\Big(\frac{|\sqrt{2}\,\sigma x-\Delta\rho|}{\lambda}\Big).
\label{SPDEapp2}
\end{equation}
It is easy to see that this equation can be obtained from Eq.~(\ref{SPDEapp1}) by the simple replacement $1/\lambda\leftrightarrow\sigma/\lambda$. Its solution in the domain of interest can then be obtained by applying this replacement to Eq.~(\ref{SXm}). This gives
\begin{equation}
S_{X^{^+}}(x,y)\approx\frac{\lambda\Delta\rho}{2\sigma\left(1-\alpha\Delta\rho\right)}\,
\ln\!\frac{1-\sqrt{2}\,\sigma x/(\lambda r_0+\Delta\rho)}{1+\sigma\Delta\rho/(\lambda r_0+\Delta\rho)}\,.
\label{SXp}
\end{equation}

\vskip0.25cm\emph{(b)~The Tunnel Exchange Coupling Integral}

It is noteworthy that both Eq.~(\ref{SXm}) and Eq.~(\ref{SXp}) are fully consistent with the result reported for $\sigma\!=\!1$ previously~\cite{BondVlad18}. The functions $\psi_{X^{^\pm}}$ one obtains by plugging these equations into Eq.~(\ref{psixy}) can be used to evaluate the tunnel exchange coupling integrals $J_{X^{^\pm}}$ in Eq.~(\ref{JXpm}). The differentiation therein can be conveniently done using the following easy-to-prove rule:
\begin{eqnarray}
\mbox{\emph{if}}~F(x,y)=F_0(x,y)e^{-A(x,y)}~~\mbox{\emph{with}}~~F_0(x,y)=Ce^{-\gamma B(x,y)},~\mbox{\emph{then}}\hskip0.75cm\nonumber\\[0.3cm]
F\frac{\partial F}{\partial(x,y)}=-\Big[\frac{\partial A}{\partial(x,y)}+\gamma\frac{\partial B}{\partial(x,y)}\Big]F^2~~\mbox{\emph{and}}~~
-\gamma\frac{\partial B}{\partial(x,y)}=\frac{1}{F_0}\frac{\partial F_0}{\partial(x,y)}\,.\nonumber
\end{eqnarray}
Here $\partial/\partial(x,y)$ stands for either $\partial/\partial x$ or $\partial/\partial y$. With this, after simplifications and elementary integration over $y$ one obtains $J_{X^{^\pm}}(\Delta\rho)$ in the form of Eq.~(2) in the main text.

Seeking the extremum for $J_{X^{^\pm}}(\Delta\rho)$ must only include the leading term in small $1/\Delta\rho$ to be consistent with the procedure of finding $S_{X^{^\pm}}$ described above. Taking the derivative of $J_{X^{^\pm}}$ over $\Delta\rho$, equating it to zero, and solving the polynomial equation obtained to the first infinitesimal order in $1/\Delta\rho$, results in $\Delta\rho_{X^{^\pm}}$ in the form of Eq.~(4) in the main text.

\vskip0.25cm\emph{(c)~Remarks on the Interlayer Coulomb Interaction Potential}

The electrostatic interaction potential energies (\ref{VC}) and (\ref{VKR}) we use in our analysis can be shown to consistently originate from the general solution to the electrostatic boundary-value problem that includes two coupled parallel monolayers. Such a solution was recently obtained by one of us (with coathors) as a byproduct in the bilayer optical probing experiment analysis (see Ref.~\cite{Lozovik19}, Appendix A). A bilayer system was considered to consist of the two parallel monolayers with individual $2D$-polarizabilities $\chi_\texttt{2D}^{\,\prime}$ and $\chi_\texttt{2D}^{\,\prime\prime}$ (in our notations) that are separated by a distance $d$ and surrounded by a dielectric medium of the static permittivity $\varepsilon$, with a point charge sitting at the origin of the cylindrical coordinate system placed in the bottom layer. In order to find the electrostatic interaction potential energy in the whole space, the Poisson's equation was solved in the Fourier space in the way similar to that reported in Ref.~\cite{Rubio11}. In the $2D$-coordinate space, the solution obtained yields the electrostatic unlike- and like-charge interaction energies of interest as follows (atomic units)
\begin{eqnarray}
V_\texttt{2D}(\rho,d)=-\!\!\int_0^\infty\!\!\!\frac{dq\,J_0(q\rho)\,e^{-qd}}{(1+qr_0^{\prime})(1+qr_0^{\prime\prime})-q^2r_0^{\prime}r_0^{\prime\prime}\,e^{-2qd}}\,,\nonumber\\
\label{BilayerSolved}\\
V_\texttt{2D}(\rho,0)=\!\int_0^\infty\!\!\!\frac{dq\,J_0(q\rho)\,[1+qr_0^{\prime\prime}(1-e^{-2qd})]}{(1+qr_0^{\prime})(1+qr_0^{\prime\prime})-q^2r_0^{\prime}r_0^{\prime\prime}\,e^{-2qd}}\,,\nonumber
\end{eqnarray}
where $r_0^\prime\!=2\pi\chi_\texttt{2D}^{\,\prime}$ and $r_0^{\,\prime\prime}\!=2\pi\chi_\texttt{2D}^{\,\prime\prime}$ are the respective screening parameters for the individual monolayers. Due to the presence of the second layer, these equations do not seem to look similar to the solitary-monolayer KR potential case. However, setting $d\!=\!\infty$ to take the top layer away makes the former zero, while the latter integrates to yield the KR potential energy (\ref{VKR}) with the effective screening length $r_0=r_0^\prime$ just as it should be.

\begin{figure}[b]
\includegraphics[scale=0.57]{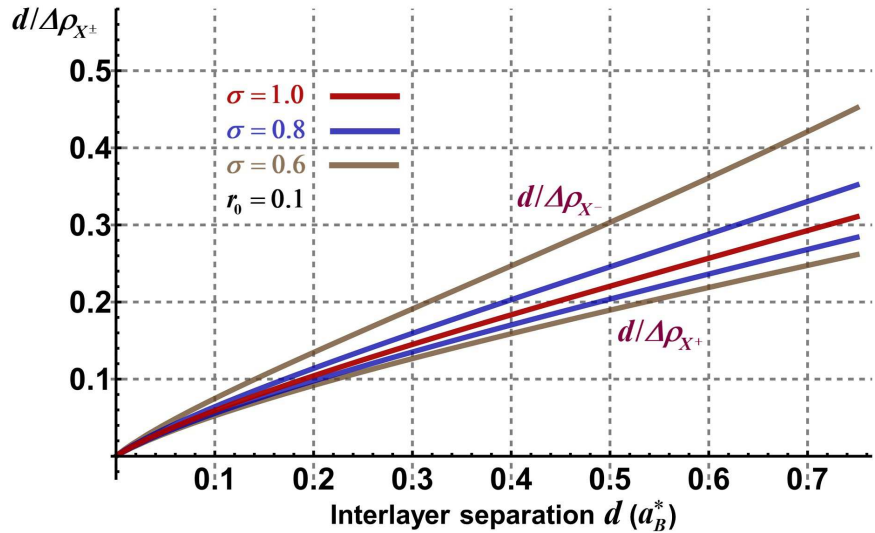}
\caption{The ratio of the interlayer distance $d$ to the equilibrium center-of-mass-to-center-of-mass separation $\Delta\rho_{X^\pm}$ of the two equivalent IEs forming the CIE, as given by Eq.~(\ref{Drho0Xpm}) for a typical set of parameters used in this work.}
\label{fig5}
\end{figure}

A close inspection of Eq.~(\ref{BilayerSolved}) reveals that due to the oscillatory behavior of the 0th order Bessel function $J_0(x)$ for all $x\!>\!1$, only $q\!\lesssim\!1/\rho$ contribute the most to the integrals there. In our case, $\rho\approx\Delta\rho_{X^\pm}$ as can be seen from Fig.~\ref{fig4}~(a). Then, in the domain $1/\Delta\rho_{X^\pm}\!<\!1$ we work within, only wave vectors $q\!\lesssim\!1/\rho\approx\!1/\Delta\rho_{X^\pm}\!<\!1$ contribute the most to both integrals in Eq.~(\ref{BilayerSolved}), so that $qd\lesssim d/\Delta\rho_{X^\pm}\!<\!d\!<\!1$ in both integrals for all $d$ we used in this work. This can also be seen from Fig.~\ref{fig5} we obtained using $\Delta\rho_{X^\pm}$ of Eq.~(\ref{Drho0Xpm}). Therefore, it is legitimate to neglect $q^2$-terms under the integrals in Eq.~(\ref{BilayerSolved}). This gives
\[
V_\texttt{2D}(\rho,d)\approx-\frac{1}{\rho}\int_0^\infty\!\!\frac{dx\,J_0(x)\,e^{-xd/\rho}}{1+x(r_0^\prime\!+\!r_0^{\prime\prime})/\rho}\,,~~~
V_\texttt{2D}(\rho,0)\approx\frac{1}{\rho}\int_0^\infty\!\!\!\frac{dx\,J_0(x)}{1+x(r_0^\prime\!+\!r_0^{\prime\prime})/\rho}\,,
\]
and the second integral turns into the KR potential energy (\ref{VKR}) with the screening length $r_0\!=\!r_0^{\prime}+r_0^{\prime\prime}$. Additionally, as per previous computational studies of monolayer TMDs~\cite{Berkelbach2013}, the monolayer screening length can be accurately represented by $c(\varepsilon_\perp\!-\!1)/2(\epsilon_1\!+\epsilon_2)$, where $c$ and $\varepsilon_\perp$ are the \emph{bulk} TMD out-of-plane translation period and in-plane dielectric permittivity, respectively. For a TMD bilayer embedded in hBN with $\varepsilon\!=5.87$ (averaged over all three directions~\cite{Laturia18}), which is the case for a variety of experiments~\cite{Lius-PKim19,BondSnoke20,Crooker19}, the typical parameters are $c\!\approx\!12\!-\!13$~\AA, $\varepsilon_\perp\!\approx\!14\!-\!17$, $\epsilon_1\!=\!(2\varepsilon_\perp\!+\varepsilon_\parallel)/3$ with $\varepsilon_\parallel\approx\varepsilon_\perp/2$~\cite{Berkelbach2013,Laturia18} and $\epsilon_2\!=\!\varepsilon$ (or vice versa), to yield $r_0\!\approx\!c(\varepsilon_\perp\!-1)/(5\varepsilon_\perp/6+\varepsilon)a_B^{\ast-1}\!\!<\!1$ as $a_B^{\ast}$ in TMDs is consistently greater than $1\,\mbox{nm}$ both by our data (see Fig.~\ref{FigTable}) and also by others~\cite{Berkelbach2013,BondSnoke20,Crooker19}. Then, we obtain $r_0/\rho\approx\!r_0/\Delta\rho_{X^\pm}\!\ll\!1$. With this in mind the denominator of the first integral above can be expanded in rapidly convergent binomial series, whereby after the term-by-term integration the interlayer electrostatic interaction energy takes the form
\[
V_\texttt{2D}(\rho,d)\approx-\frac{1}{\rho}\int_0^\infty\!\frac{dxJ_0(x)\,e^{-xd/\rho}}{1+xr_0/\rho}\approx-\frac{1}{\sqrt{\rho^2+d^2}}\,
\Big(1-\frac{d}{\rho}\frac{1}{1+d^2/\rho^2}\frac{r_0}{\rho}+\cdots\Big)\,.
\]
Here, the second term in parentheses comes out as the 2nd (not the 1st as one would expect!) order of smallness since $d/\rho\approx d/\Delta\rho_{X^\pm}\!<\!1$ as demonstrated in Fig.~\ref{fig5}, and so it can be safely dropped along with the rest of higher infinitesimal order terms, whereby one arrives at the interlayer Coulomb interaction (\ref{VC}) we used in our calculations throughout this work. Note also that, even more generally, this series expansion can be seen to be uniformly suitable for all $\rho\!\ge\!0$, including $\rho\!\sim\!0$ as well, in which case the second term in parentheses comes out as the 1st order of smallness in $r_0/d$ and \emph{still} can be dropped for $d$ large enough, whereby one \emph{still} arrives at Eq.~(\ref{VC}) --- now in the \emph{classical} electrostatic Coulomb interaction regime of two space-separated point charges with intercharge distance written in cylindrical coordinates.

\begin{figure}[b]
\includegraphics[scale=0.5]{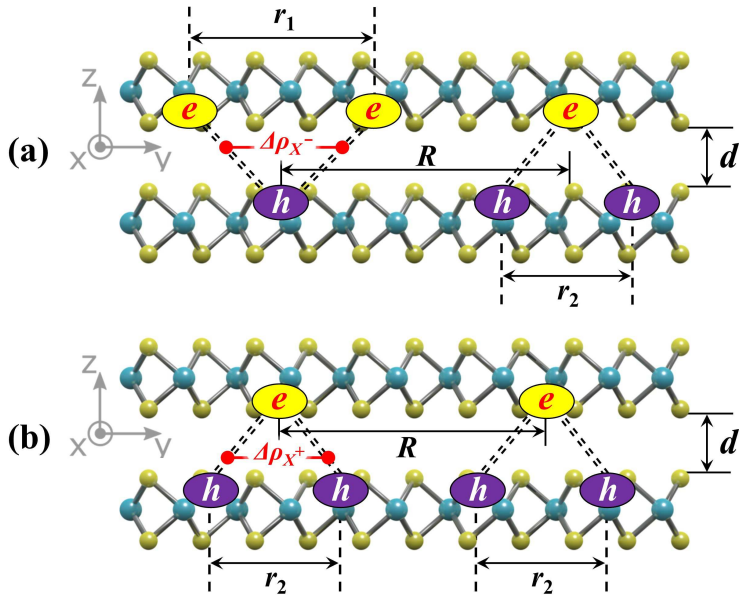}
\caption{The coplanar pairwise interaction geometry for the unlike-charge~(a) and like-charge~(b) IE complexes (trions) in a TMD bilayer.}
\label{fig6}
\end{figure}

\subsection{The pairwise interaction potentials for charged interlayer excitons}

As can be seen from the two special cases shown in Fig.~\ref{fig6}~(a) and (b), the long-range Coulomb interaction of the pair of CIEs (trions) depends on the relative orientation of the triangles formed by the three charges in a trion complex. The exact interaction potential includes nine terms to couple the electrons and holes in the two spatially separated complexes. To simulate the actual potential energy surfaces we use the Coulomb interaction coupling of Eq.~(\ref{VC}) for the (unlike) charges located in the distinct monolayers and the KR interaction coupling of Eq.~(\ref{VKR}) for the (like) charges confined to the same monolayer. The explicit coupling parameter dependence is given by the functions $V_{\texttt{KR}}(R,r_1,r_2,r_0)$ and $V_{\texttt{C}}(R,r_1,r_2,d)$ specified below, where $R$ is the trion-trion center-of-mass-to-center-of-mass distance and $r_{1,2}$ are the distances between the like charges in the first and second trion of the interacting trion pair. In general, $r_1\!\ne\!r_2$ for the unlike-charge trion-trion coupling and $r_1\!=\!r_2$ for the like-charge trion-trion coupling as sketched in Fig.~\ref{fig6}~(a) and (b). Using the standard triangle similarity theorems, these distances come out as $\lambda\Delta\rho_{X^{^+}}$ and $(\lambda/\sigma)\Delta\rho_{X^{^-}}$ for the positive and negative trion, respectively. This is why for unlike-charge trion pairs, $r_1$ can only be equal to $r_2$ if $\sigma\!=\!1$ (or $m_e\!=\!m_h$).

\begin{figure}[b]
\includegraphics[scale=0.5]{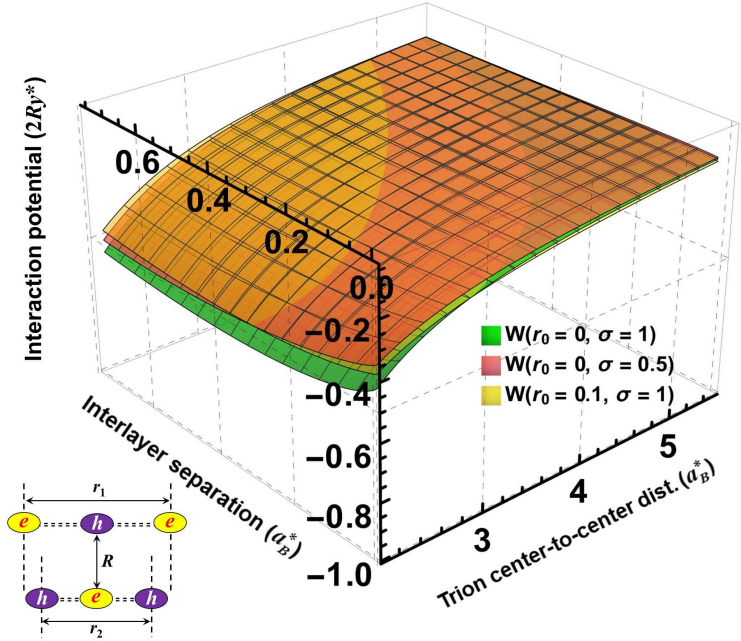}
\caption{The pairwise interaction potential energy $W$ as given by Eq.~(\ref{W}) for the two unlike-charge IEs in the parallel biplanar relative orientation sketched at bottom left (top view).}
\label{fig7}
\end{figure}

\vskip0.25cm\emph{(a)~The Pairwise Interaction Potentials for Unlike-Charge IEs}

In this case, two most likely relative orientations are supported by symmetry for a pair of triangle-shaped complexes in bilayer structures we deal with. They are the coplanar and parallel biplanar orientation. Their side and top views are shown in Fig.~\ref{fig6}~(a) and in the bottom-left inset of Fig.~\ref{fig7}, respectively. For the former, counting $e$-$h$ couplings in Fig.~\ref{fig6}~(a) counterclockwise from top left, the total interaction potential energy $U$ takes the form
\begin{eqnarray}
U=u_{1}+u_{2}+u_{3},
\label{U}
\end{eqnarray}
\[
u_{1}=V_\texttt{KR}(R+r_{1}/2)
+V_\texttt{C}\Big\{\!\sqrt{\big[R+(r_{1}-r_{2})/2\big]^{2}+d^{2}}\Big\}
+V_\texttt{C}\Big\{\!\sqrt{\big[R+(r_{1}+r_{2})/2\big]^{2}+d^{2}}\Big\},
\]
\[
u_{2}=V_\texttt{KR}(R-r_{2}/2)+V_\texttt{KR}(R+r_{2}/2)+V_\texttt{C}\Big(\!\sqrt{R^{2}+d^{2}}\,\Big),
\]
\[
u_{3}=V_\texttt{KR}(R-r_{1}/2)
+V_\texttt{C}\Big\{\!\sqrt{\big[R-(r_{1}+r_{2})/2\big]^{2}+d^{2}}\Big\}
+V_\texttt{C}\Big\{\!\sqrt{\big[R-(r_{1}-r_{2})/2\big]^{2}+d^{2}}\Big\}.
\]
For the latter, from the inset in Fig.~\ref{fig7} the total interaction potential $W$ comes out as
\begin{eqnarray}
W=2w_{1}+w_{2},
\label{W}
\end{eqnarray}
\[
w_{1}=V_\texttt{C}\Big[\!\sqrt{(r_{1}-r_{2})^{2}/4+R^{2}+d^{2}}\Big]
+V_\texttt{KR}\Big(\!\sqrt{r_{1}^{2}/4+R^{2}}\,\Big)
+V_\texttt{C}\Big[\!\sqrt{(r_{1}+r_{2})^{2}/4+R^{2}+d^{2}}\Big],
\]
\[
w_{2}=2V_\texttt{KR}\Big(\!\sqrt{r_{2}^{2}/4+R^{2}}\,\Big)+V_\texttt{C}\Big(\!\sqrt{R^{2}+d^{2}}\,\Big).
\]

The calculated interaction potentials $U$ and $W$ are presented in Fig.~3~(a) of the main text and in Fig.~\ref{fig7} herewith, respectively. The former is seen to be over an order of magnitude more attractive than the latter in the same parameter range, which is why the $W$ interaction potential energy is neglected in the analysis we report about in the main text.

\vskip0.25cm\emph{(b)~The Pairwise Interaction Potentials for Like-Charge IEs}

In this case, both coplanar and parallel biplanar relative orientations of the triangle-shaped complexes are strongly repulsive and, in general, are different for positively and negatively charged trion pairs. The side view of the coplanar orientation of two positive trions is shown in Fig.~\ref{fig6}~(b). The top view of their parallel biplanar orientation can be obtained from the sketch in Fig.~\ref{fig7} by setting $r_1\!=\!r_2$ and relabeling $e\!\leftrightarrow\!h$ in one of the trions. For the former, counting $e$-$h$ couplings in Fig.~\ref{fig6}~(b) counterclockwise from top left, the total interaction potential energy $V$ takes the form
\begin{eqnarray}
V=v_{1}+v_{2}+v_{3},
\label{V}
\end{eqnarray}
\[
v_{1}=V_\texttt{KR}(R)+V_\texttt{C}\Big[\!\sqrt{(R-r_{2}/2)^{2}+d^{2}}\Big]+V_\texttt{C}\Big[\!\sqrt{(R+r_{2}/2)^{2}+d^{2}}\Big],
\]
\[
v_{2}=V_\texttt{KR}(R)+V_\texttt{KR}(R+r_{2})+V_\texttt{C}\Big[\!\sqrt{(R+r_{2}/2)^{2}+d^{2}}\Big],
\]
\[
v_{3}=V_\texttt{KR}(R-r_{2})+V_\texttt{KR}(R)+V_\texttt{C}\Big[\!\sqrt{(R-r_{2}/2)^{2}+d^{2}}\Big].
\]
For the latter, the total interaction potential energy $\bar{V}$ can be obtained from Eq.~(\ref{W}) by setting $r_1\!=\!r_2$ and simultaneously swapping $V_\texttt{KR}\!\leftrightarrow\!V_\texttt{C}$ and $R^2\!\leftrightarrow\!R^2+d^2$. This gives
\begin{eqnarray}
\bar{V}=2\bar{v}_{1}+\bar{v}_{2},
\label{Vbar}
\end{eqnarray}
\[
\bar{v}_{1}=V_\texttt{KR}(R)+V_\texttt{C}\Big(\!\sqrt{r_{2}^{2}/4+R^{2}+d^{2}}\,\Big)+V_\texttt{KR}\Big(\!\sqrt{r_{2}^{2}+R^{2}}\,\Big),
\]
\[
\bar{v}_{2}=2V_\texttt{C}\Big(\!\sqrt{r_{2}^{2}/4+R^{2}+d^{2}}\,\Big)+V_\texttt{KR}(R).
\]
For a negatively charged trion pair, $r_2$ should be replaced with $r_1$ in both of these equations.

A close inspection of Eqs.~(\ref{V}) and (\ref{Vbar}) reveals their very similar repulsive behavior and in fact their coincidence when $R$ is greatly different from $r_2$ (both greater and less than). The calculated interaction potential $V$ of Eq.~(\ref{V}) is presented in Fig.~3~(a) of the main text.

\subsection{Like-charge trion Wigner crystallization parameters}

An ensemble of repulsively interacting particles (or quasi\-particles, structureless or compound) forms a~Wig\-ner lattice when its average potential interaction energy exceeds average kinetic energy, $\langle V\rangle/\langle K\rangle\!=\!\Gamma_0\!>\!1$ (see, e.g., Ref.~\cite{Platzman74}). For like-charge trions in Fig.~\ref{fig6}~(b), the Coulomb repulsion at large $R$ ($\gg\!r_2$) is strengthened at shorter $R$ by the dipole-dipole repulsion of their collinear permanent dipole moments directed perpendicular to the hetero\-structure plane. These are the two major terms of the power series expansion in $r_2/R\;(<\!1)$ of the repulsive pairwise interaction potential $V$ presented in Fig.~3~(a) of the main text. With rotational kinetic energy neglected for the reasons explained in the main text, the like-charge trion critical density $n_{cX^{\pm}}$ and temperature $T^\texttt{(W)}_{cX^{\pm}}$ can be obtained by drawing an analogy to the 2D electron gas system~\cite{Platzman74} to include the extra dipole-dipole repulsion term.

\vskip0.25cm\emph{(a)~The Critical Density}

With the commonly used notations preserved, we go on with using the atomic units introduced previously. For trion-trion separation distances $R$ greater than the size of the trion ($R\!\gg\!r_{1,2}$), the first order power series expansion of the average repulsive trion-trion interaction potential takes the form
\begin{equation}
\langle V\rangle=\frac{1}{R}\Big(1+\frac{d^2}{R^2}\Big)=\sqrt{\pi n}\,\big(1+d^2\pi n\big),
\label{Vav}
\end{equation}
where $n\!=\!1/\pi R^2$ is the trion surface density. Our trions are compound fermions with the occupation number
\begin{equation}
n_\mathbf{k}=\frac{1}{e^{\,\beta(E_\mathbf{k}-\bar{\mu})}+1}\,,
\label{occupnumber}
\end{equation}
where $\beta\!=\!1/k_BT$, $E_\mathbf{k}\!=\!\hbar^2k^2/2M$, $M\!=\!M_{X^{\pm}}$ and $\bar{\mu}$ being the trion total mass and chemical potential, respectively. At zero $T$ this turns into a unit-step function to give $n$ in Eq.~(\ref{Vav}) in the form
\begin{equation}
n=\frac{\langle N\rangle}{S}=\frac{2}{S}\sum_\mathbf{k}n_\mathbf{k}=\frac{2}{S}\frac{S}{(2\pi)^2}\,2\pi\!\!\int_0^{k_F}\!\!\!dk\,k=\frac{k_F^2}{2\pi}\,,
\label{n}
\end{equation}
where $S$ is the surface area and $k_F$ is the trion Fermi-momentum. The average kinetic energy per particle can then be written as
\begin{equation}
\langle K\rangle=\frac{2}{\langle N\rangle}\sum_\mathbf{k}E_\mathbf{k}n_\mathbf{k}
=\frac{2}{\langle N\rangle}\frac{S}{(2\pi)^2}\,2\pi\!\!\int_0^{k_F}\!\!\!dk\,k\,\frac{\hbar^2k^2}{2M}
=\frac{\pi S}{\langle N\rangle}\frac{\hbar^2}{2M}\Big(\frac{k_F^2}{2\pi}\Big)^2=\frac{\hbar^2}{2M}\,\pi n
\label{Kav}
\end{equation}
to result, with $\langle V\rangle$ of Eq.~(\ref{Vav}), in
\begin{equation}
\Gamma_0=\frac{\langle V\rangle}{\langle K\rangle}=\frac{2M}{\hbar^2}\,\frac{1+d^2\pi n}{\sqrt{\pi n}}=\frac{2}{g}\,\frac{1+d^2\pi n}{\sqrt{\pi n}}\,,
\label{G0}
\end{equation}
where $g$ stands for the ratio of the electron-hole reduced mass to the trion total mass
\begin{equation}
g=\frac{\mu}{M}=\frac{\mu}{M_{X^{\pm}}}=g_{\pm}(\sigma)=\Big(3+\Big\{\!\begin{array}{c}1\\[-0.35cm]2\end{array}\!\Big\}\,\sigma
+\Big\{\!\begin{array}{c}2\\[-0.35cm]1\end{array}\!\Big\}\,\frac{1}{\sigma}\Big)^{\!\!-1}.
\label{g}
\end{equation}

Introducing the new variable $t\!=\!d\sqrt{\pi n}$ turns Eq.~(\ref{G0}) into a quadratic equation
\[
t^2-\frac{g\Gamma_0}{2d}\,t+1=0
\]
with two roots as follows
\[
t_{1,2}=\frac{g\Gamma_0}{4d}\pm\sqrt{\Big(\frac{g\Gamma_0}{4d}\Big)^{\!2}-1}\,,
\]
of which only one, $t_2$, stays finite as $d$ goes down to zero. This root leads to
\begin{equation}
n_{cX^{\pm}}\!=\!\frac{2}{\pi d^2}\Big(\frac{g_\pm\Gamma_{0}}{4d}\Big)^{\!2}\Big[
1-\frac{1}{2}\Big(\frac{4d}{g_\pm\Gamma_{0}}\Big)^{\!2}-\sqrt{1-\Big(\frac{4d}{g_\pm\Gamma_{0}}\Big)^{\!2}}\,\Big]
\label{nc}
\end{equation}
and reproduces the result of Ref.~\cite{Platzman74} for $d\rightarrow0$ and $g_\pm\!=\!1$.

\vskip0.25cm\emph{(b)~The Critical Temperature}

For arbitrary nonzero $T$, using Eq.~(\ref{occupnumber}) with the new variable $x\!=\!\hbar k\sqrt{\beta/2M}$, the trion surface density (\ref{n}) can be written in a parametric form as follows
\begin{equation}
n=\frac{\langle N\rangle}{S}=\frac{2}{S}\frac{S}{(2\pi)^2}\,2\pi\!\!\int_0^\infty\!\!\!\frac{dk\,k}{e^{\,\beta(E_\mathbf{k}-\bar{\mu})}+1}
=\frac{2M}{\hbar^2\pi\beta}\!\int_0^\infty\!\!\!dx\,x\frac{ze^{-x^2}}{1+ze^{-x^2}}\,,\;\;\;z=e^{\beta\bar{\mu}}\ge0\,.
\label{nT}
\end{equation}
Similarly, the average kinetic energy per particle of Eq.~(\ref{Kav}) takes the form
\begin{equation}
\langle K\rangle=\frac{2}{\langle N\rangle}\frac{S}{(2\pi)^2}\,2\pi\frac{\hbar^2}{2M}\!\int_0^\infty\!\!\!\frac{dk\,k^3}{e^{\,\beta(E_\mathbf{k}-\bar{\mu})}+1}
=\frac{1}{\beta}\,\frac{\displaystyle\int_0^\infty\!\!\!dx\,x^3\frac{ze^{-x^2}}{1+ze^{-x^2}}}{\displaystyle\int_0^\infty\!\!\!dx\,x\frac{ze^{-x^2}}{1+ze^{-x^2}}}\,.
\label{KavT}
\end{equation}
After the power series expansions of their respective denominators, these integrals can further be represented in terms of the gamma and polylogarithm functions following the rule
\begin{equation}
\sum_{m=1}^\infty\int_0^\infty\!\!\!x^n\big(\!-\!ze^{-x^2}\big)^mdx=\frac{\Gamma\big[(n+1)/2\big]}{2}\,\mbox{Li}_{(n+1)/2}(-z)
=\frac{\Gamma\big[(n+1)/2\big]}{2}\sum_{m=1}^\infty\frac{(-z)^m}{m^{(n+1)/2}}\,.
\label{rule}
\end{equation}
In the classical limit (high $T$ and/or low density; see, e.g., Ref.~\cite{Chandler}), one has $e^{\,\beta(E_\mathbf{k}\!-\bar{\mu})}\!\gg\!1$, so that the occupation number (\ref{occupnumber}) takes the form $n_\mathbf{k}\!=\!e^{-\beta(E_\mathbf{k}-\mu)}\!=ze^{-\beta E_\mathbf{k}}$ to simplify $n$ in Eq.~(\ref{nT}) as follows
\begin{equation}
n=\frac{2Mz}{\hbar^2\pi\beta}\!\int_0^\infty\!\!\!dx\,xe^{-x^2}\!=\frac{Mz}{\hbar^2\pi\beta}\,,
\label{nclass}
\end{equation}
whereby the kinetic energy per particle of Eq.~(\ref{KavT}) takes the form
\begin{equation}
\langle K\rangle=\frac{2Mz}{\hbar^2\pi\beta^2n}\int_0^\infty\!\!\!dx\,x^3e^{-x^2}\!=\frac{Mz}{\hbar^2\pi\beta^2n}=\frac{1}{\beta}=k_BT
\label{Kavclass}
\end{equation}
as expected from the energy equipartition theorem of classical statistical mechanics.

Plugging Eqs.~(\ref{Vav}) and (\ref{Kavclass}) in Eq.~(\ref{G0}) gives the equality $\,\pi n\big(1+d^2\pi n\big)^2\!=\big(\Gamma_0k_BT\big)^2$. In this equation, to make it consistent with the approximation Eq.~(\ref{Vav}) is valid within, one has to discard the terms with powers of $d$ higher than $d^2$. The quadratic equation thus obtained gives two roots for $n(T)$, one of which is manifestly negative and so to be discarded. Equating the other root to $n_{cX^{\pm}}$ of Eq.~(\ref{nc}) gives the constraint for the critical temperature. Solving it for $T$ subject to keeping powers of $d$ no greater than $d^2$, leads to
\[
k_{B}T^\texttt{(W)}_{cX^{\pm}}=\frac{4}{g_{\pm}\Gamma_0^2}
\]
(in the units of $Ry^\ast$) with $g_{\pm}(\sigma)$ given by Eq.~(\ref{g}). This agrees with Ref.~\cite{Platzman74} for $g_\pm\!=\!1$.

\section{Acknowledgments} This research is supported by the U.S. Department of Energy, Office of Science, Office of BES under award No.$\,$DE-SC0007117 (I.V.B.), by the U.S. ARO grant No.$\,$W911NF1810433 (O.L.B., R.Ya.K.), and by the RFBR grants No.$\,$20-02-00410 and No.$\,$20-52-00035 (Y.E.L.).

\section{Author contributions}

I.V.B. conceived the project, developed the theory, carried out theoretical and numerical calculations, and wrote the final version of the manuscript. O.L.B. and R.Ya.K. contributed in-depth assessments of exciton and trion interaction potentials. Y.E.L. provided expertise in exciton many-particle correlations and crystallization phenomena. All authors discussed the results and commented on the ways to best represent them in the manuscript.

\section{Competing interests}

The authors declare no competing financial interests.

\end{document}